\documentclass[12pt]{article}
\pdfoutput=1

\usepackage{setspace}
\usepackage{amsmath}
\usepackage{amssymb}
\usepackage{colortbl}
\usepackage[table,x11names]{xcolor}
\usepackage{graphicx}
\usepackage{hyperref}
\usepackage{cite}
\usepackage{slashed}
\usepackage{pifont}
\usepackage{longtable}
\usepackage{multirow}
\usepackage{spreadtab}
\usepackage{ltablex}

\usepackage{booktabs}

\definecolor{nicered}{rgb}{0.6,0,0}
\definecolor{nicegreen}{rgb}{0.1,0.5,0.1}
\definecolor{niceblue}{rgb}{0,0.4,0.8}
\hypersetup{colorlinks,citecolor= nicegreen,linkcolor=
  nicered,urlcolor=niceblue}
\allowdisplaybreaks

\begin{document}
\begin{titlepage}

  \newcommand{\AddrLiege}{{\sl \small IFPA, D\'ep. AGO, Universit\'e de
      Li\`ege, B\^at B5, Sart Tilman B-4000 Li\`ege 1, Belgium}}
  \newcommand{\AddrUFSM}{{\sl \small  Universidad T\'ecnica 
      Federico Santa Mar\'{i}a - Departamento de F\'{i}sica\\
      Casilla 110-V, Avda. Espa\~na 1680, Valpara\'{i}so, Chile}}
  \vspace*{0.5cm}
\begin{center}
  \textbf{\Large Radiative accidental matter}
  \\[9mm]
  D. Aristizabal Sierra$^{a,b,}$\footnote{e-mail address: {\tt
      daristizabal@ulg.ac.be}},
  C. Simoes$^{a,}$\footnote{email address: {\tt csimoes@ulg.ac.be}},
  D. Wegman$^{a,}$\footnote{email address: {\tt dwegman@ulg.ac.be}}
  \vspace{0.8cm}\\
  $^a$\AddrLiege\\[5mm]
  $^b$\AddrUFSM\\[5mm]
\end{center}
\vspace*{0.2cm}
\begin{abstract}
  \onehalfspacing
  Accidental matter models are scenarios where the beyond-the-standard
  model physics preserves all the standard model accidental and
  approximate symmetries up to a cutoff scale related with lepton
  number violation. We study such scenarios assuming that the new
  physics plays an active role in neutrino mass generation, and show
  that this unavoidably leads to radiatively induced neutrino
  masses. We systematically classify all possible models and determine
  their viability by studying electroweak precision data, big bang
  nucleosynthesis and electroweak perturbativity, finding that the
  latter places the most stringent constraints on the mass
  spectra. These results allow the identification of minimal radiative
  accidental matter models for which perturbativity is lost at high
  scales. We calculate radiative charged-lepton flavor violating
  processes in these setups, and show that $\mu\to e \gamma$ has a
  rate well within MEG sensitivity provided the lepton-number
  violating scale is at or below $10^6\,$~GeV, a value (naturally)
  assured by the radiative suppression mechanism. Sizeable
  $\tau\to \mu \gamma$ branching fractions within SuperKEKB
  sensitivity are possible for lower lepton-number breaking scales. We
  thus point out that these scenarios can be tested not only in direct
  searches but also in lepton flavor-violating experiments.
\end{abstract}
\end{titlepage}
\setcounter{footnote}{0}

\section{Introduction}
\label{sec:intro}
Various theoretical and experimental arguments support the idea that
at certain energy scale new degrees of freedom should be operative. A
solution to the electroweak (EW) hierarchy problem demands this scale
to be order TeV and leads to new physics potentially testable at the
LHC. This new physics is expected to address not only the hierarchy
problem, but to account as well for experimental-driven puzzles such
as the origin of dark matter \cite{Ade:2015xua}, neutrino masses
\cite{Fukuda:1998mi,Ahmad:2002jz,Eguchi:2002dm} and the baryon
asymmetry of the universe \cite{Ade:2015xua}. Typical models follow a
rather simple approach in which sectors subject to strong
phenomenological constraints are decoupled, while those for which
experimental bounds (of whatever nature) are somewhat weaker are
associated with low-energy scales.

This is the case for lepton number-violating (LNV) physics, which in
the absence of a ``non-conventional'' suppression
mechanism\footnote{By non-conventional we refer to mechanisms where
  the suppression does not rely on the presence of GUT-scale states,
  i.e. radiative or slightly broken lepton number mechanisms, see
  sec. \ref{sec:acc-mat-rep}.} demands GUT-scale physics. If one adopts
that approach, the smallness of neutrino masses ``naturally'' arise
due to the presence of decoupled states that can be related as well
with a solution to the baryon asymmetry problem (see
e.g. \cite{Davidson:2008bu,Fong:2013wr,Hambye:2012fh,AristizabalSierra:2010mv,
  Sierra:2014tqa}) and that allows for embeddings of such scenarios in
GUTs. In such picture the LNV physics, apart from accounting for
low-energy neutrino observables, does not leave any experimental
trace, something consistent with direct and indirect experimental
searches, see e.g.
\cite{ATLAS:2012ak,Khachatryan:2014dka,Adam:2013mnn}. The TeV sector
that accounts for the hierarchy problem and that might involve dark
matter (DM) states, contribute to various rare processes whose current
bounds are as well tight
\cite{Agashe:2014kda,Isidori:2010kg}. Consistency therefore requires a
mechanism that allows for TeV states while explaining the absence of
signals in the high-intensity frontier.

A rather popular approach to such problem is given by the minimal
flavor violation hypothesis, that postulates that the only source of
flavor violation are the standard model (SM) Yukawa couplings,
regardless of the nature of the new physics
\cite{D'Ambrosio:2002ex}. In the quark sector this assumption
automatically reconciles TeV-scale physics with the non-observation of
flavor-violating signals in indirect searches. In the lepton sector
despite not being univocally implementable it leads to a consistent
picture that in its minimal realizations entails quite a few
predictions for lepton flavor-violating processes
\cite{Alonso:2011jd}. Accidental matter, is another interesting
approach that although does not aim at addressing the EW hierarchy
problem it enables for new order-TeV physics without invoking any
special flavor structure \cite{DiLuzio:2015oha}. The idea is that the
new physics should preserve all the SM exact and approximate
symmetries up to a cutoff scale $\Lambda$, above which these
symmetries are presumably broken by a larger theory. Assuming that
this cutoff scale is universal and determined by the constraints
implied by lepton number violation (in the absence of a
``non-conventional'' suppression mechanism), a simple but compelling
\textit{minimal} picture emerges with new order-TeV states that do not
conflict with negative results from indirect experimental searches.

In this paper we define and study alternative forms of accidental
matter models which we dub \textit{radiative accidental
  matter}\footnote{A systematic classification of $U(1)_{B-L}$
  loop-induced neutrino mass models has been presented in
  ref. \cite{Ho:2016aye}.}. For that aim we allow for a mismatch among
the different scales involved: LNV, quark- and lepton-flavor violating
(QFV and LFV), baryon-number violating (BNV) and the scale at which
perturbativity is lost. Furthermore, we show that if the accidental
matter representation plays an active role in neutrino mass generation
this unavoidably leads to radiative neutrino masses (hence the name
radiative accidental matter), thus providing a ``natural'' suppression
mechanism that enables for low-scale lepton number violation, which
reopens the possibility of potentially large LFV effects, thus
increasing the testability of these scenarios.

The full set of accidental matter representations are derived from the
condition that the new degrees of freedom preserve the SM accidental
and approximate symmetries up to a certain cutoff scale. This set
contains lower- and higher-order $SU(2)$ representations, with the
latter in some cases defining minimal DM models
\cite{Cirelli:2005uq,Cirelli:2007xd,Cirelli:2009uv}. This however is
not the case in the scenarios we will discuss. As has been recently
pointed out in refs. \cite{Sierra:2016qfa,Ahriche:2016rgf}, once a
lower cutoff scale is allowed (in this case related with neutrino
physics constraints) the neutral component contained in the
representation decays fast, with typical lifetimes amounting to
$\mu$s. This observation combined with a lower perturbative scale
enables non-vanishing hypercharge sextets, that otherwise would be
forbidden \cite{DiLuzio:2015oha}. Moreover, in our analysis we will
not consider hypercharge-zero septets since their quartic scalar
couplings reach Landau poles at rather low scales, $\sim 10^7\,$~GeV
\cite{Hamada:2015bra}.

The rest of this paper is organized as follows. In
sec. \ref{sec:acc-mat-rep} we discuss the different energy scales of
the problem, define a benchmark scenario for radiative accidental
matter and determine the set of relevant representations. In
sec. \ref{sec:unstability} we present our arguments for cosmological
instability of higher-order $SU(2)$ representations, paying special
attention to the scalar sextet. In
sec. \ref{sec:additional-constraints} various phenomenological
constraints are discussed, while in sec. \ref{sec:uv-completions} we
examine the different UV completed radiative accidental matter models
and study their perturbative behavior. In sec. \ref{sec:LFV} we study
in a fairly model-independent way LFV processes in these
scenarios. Finally, in sec. \ref{sec:concl} we summarize and present
our conclusions. In app.~\ref{sec:explicit-UV-completions} we present
the two-loop renormalization group equations (RGEs) that we have used
in our analysis.
\section{Effective scales in accidental matter scenarios}
Accidental matter models are weak-scale extensions of the SM in which
the beyond-SM (BSM) degrees of freedom ($\boldsymbol{R}$) preserve the
accidental and approximate symmetries of the SM at the renormalizable
level. Thus, this means that even if present at the renormalizable
level their effects will not be accessible in indirect searches. One
might wonder whether departures from their standard formulation could
change that picture. For that aim one can consider the SM as the
renormalizable part of a larger Lagrangian:
\begin{equation}
  \label{eq:lag-full}
  \mathcal{L}=\mathcal{L}_\text{SM}^\text{Ren} + 
  \sum_{N>4}\frac{\mathcal{C}_N}{\Lambda^{N-4}_\text{Eff}}\mathcal{O}_N\, ,
\end{equation}
where $\mathcal{L}_\text{SM}^\text{Ren}$ are the renormalizable SM
interactions, whereas the second term are effective operators where
the only dynamical degrees of freedom are SM fields. They do break the
SM accidental and approximate symmetries and so their effects include
CP and flavor violation in the quark and lepton sectors as well as
baryon and lepton number breaking.

In the standard approach the coefficients of the effective expansion
$\mathcal{C}_N$ are assumed to be $\mathcal{O}(1)$. When combined with
the assumption that neutrinos are Majorana particles, this fixes the
effective scale, $\Lambda_\text{Eff}\sim 10^{15}\,$~GeV.
Automatically then, all possible signatures related with departures
from SM accidental and approximate symmetries are suppressed, thus
explaining their absence in indirect searches. Testability of these
scenarios is possible only if the new states can be directly produced
and detected in collider experiments. Otherwise the new physics,
although present, could be hard---if not impossible---to reveal.  A
possible departure from the standard formulation consists then in
allowing for lower cutoff scales that in turn allow for other
observables to have sizeable values, thus increasing the testability
of these scenarios.

\begin{figure}
  \centering
  \includegraphics[scale=1]{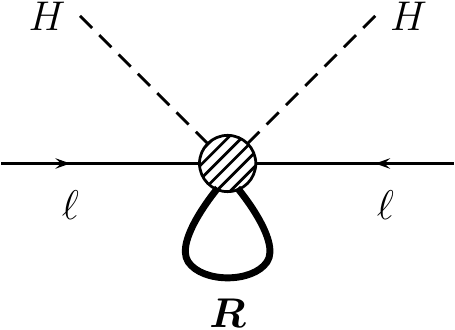}
  \caption{Dimension five LNV effective operator in the limit SM +
    $\boldsymbol{R}$. The presence of $\boldsymbol{R}$ allows for a
    \textit{naturally} small expansion coefficient $\mathcal{C}$ and
    therefore for a lower effective scale.}
  \label{fig:dimension-five-LNV-DM}
\end{figure}

The effective operators in (\ref{eq:lag-full}) are subject to
different phenomenological constraints, with the most stringent limits
enforced by neutrino masses on the LNV ones. Without any further
assumption the leading-order LNV operator is given by the dimension
five Weinberg operator
\begin{equation}
  \label{eq:weinberg}
  \frac{\mathcal{C}_{ij}}{\Lambda}\,\left(\overline{\ell^c_i}\,i\tau_2 H\,\right)
  \left(H^T\,i\tau_2\,\ell_j\right)\, ,
\end{equation}
that after EW symmetry breaking leads to
$m_\nu\sim \mathcal{C}\,v^2/\Lambda$. A lower cutoff scale, say
$\mathcal{O}(\text{TeV})$, is possible provided
$\mathcal{C}\sim 10^{-12}-10^{-11}$. There are two generic mechanisms
through which such a small coefficient can be \textit{naturally}
obtained: (i) the operator is related with a slightly broken symmetry
(for example slightly broken lepton number)
\cite{Mohapatra:1986bd,Gavela:2009cd,Sierra:2012yy,Dev:2012sg}, (ii) the operator
is radiatively induced. Though mechanisms of type (i) can be
envisaged, in the presence of SM + $\boldsymbol{R}$ the natural option
relies on (ii). Let us discuss this in more detail. Below $\Lambda$
the only degrees of freedom are the SM fields and $\boldsymbol{R}$,
and so lepton number violation should be determined by effective
operators (as required by the accidental matter scenario). Since
$\boldsymbol{R}\otimes \boldsymbol{R}\supset \boldsymbol{1} \oplus
\boldsymbol{3} \oplus \boldsymbol{5} \oplus \dots \oplus
(2\boldsymbol{R}-1)$,
the effective operator in (\ref{eq:weinberg}) can be endowed with the
new degree of freedom as shown in
fig. \ref{fig:dimension-five-LNV-DM}. Therefore, in this case the
effective expansion coefficient is suppressed by the loop factor and
by extra couplings that originate in the UV completed theory. Roughly
it can be written as $\mathcal{C}\sim Y^4_\nu/16\pi^2$, which means
that $\Lambda$ can be as small as $\sim 10^5\,$~GeV for
$Y_\nu\sim h_\tau$ ($\tau$ Yukawa coupling). These values mean that
the degrees of freedom of the UV theory, although not reachable at the
LHC can manifest in indirect searches, e.g. in $\mu \to e \gamma$,
$\mu\to 3e$ or in $\mu-e$ conversion in nuclei, processes that
are/will be searched for in MEG \cite{Adam:2013mnn}, $Mu3e$
\cite{Berger:2014vba,Bravar:2015vja} and PRISM/PRIME
\cite{Kuno:2005mm,Barlow:2011zza} respectively (see
e.g. refs. \cite{Sierra:2012yy,Alonso:2012ji} for phenomenological
studies).

\begin{figure}
  \centering
  \includegraphics[scale=0.7]{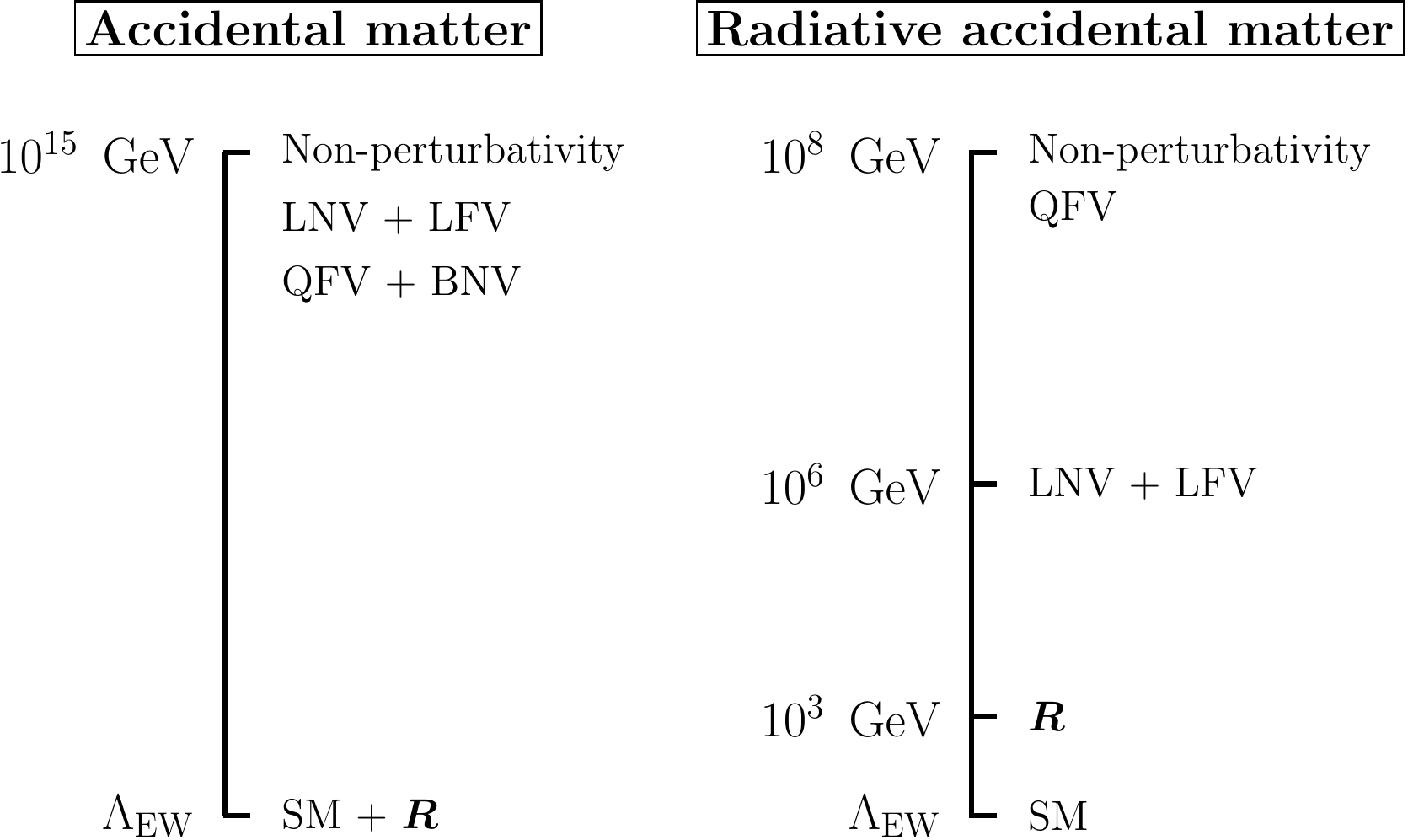}
  \caption{Energy scales in \textit{standard} (left-hand side) and
    radiative (right-hand side) accidental matter models (benchmark
    scenario). Note that the LNV, LFV, and QFV scales in the latter
    are smaller, thus allowing for potential observability of the
    corresponding processes. The energy scales on the right side
    represent the benchmark scenario we will use for our discussion.}
  \label{fig:scales}
\end{figure}

Some words are in order regarding the effective scale for the operator
in (\ref{eq:weinberg}). This scale determines the cutoff scale where
different UV completions---involving new states---enable writing down
the operator in fig. \ref{fig:dimension-five-LNV-DM} through
renormalizable couplings. This scale differs from that where
perturbativity is lost ($\Lambda_\text{Landau-pole}$), in contrast to
standard accidental matter scenarios where these two scales match (see
fig. \ref{fig:scales}). At scales above $\Lambda$ new states kick in
(generically denoted by $\boldsymbol{R}^\prime$), and their
renormalizable interactions break lepton number and lepton flavor, but
quark flavor and baryon number are still symmetries of the
renormalizable Lagrangian (at that scale). Since the new states
contribute to $\alpha_1$ and $\alpha_2$ ($\alpha_i=g_i^2/4\pi$)
running, perturbativity is lost more rapidly than in the case SM +
$\boldsymbol{R}$. The exact value where this happens depends upon the
number and dimensionality of the new representations. Assuming that
perturbativity is restored by a larger gauge theory, one could expect
quark flavor and/or baryon number to be broken at that scale. If only
quark flavor is broken, $\Lambda_\text{Landau-pole}\sim 10^8\,$~GeV
suffices to satisfy current constraints on QFV processes
\cite{Cirigliano:2013lpa}, otherwise
$\Lambda_\text{Landau-pole}\gtrsim 10^{15}\,$~GeV is required to
ensure proton stability. Here we will select viable accidental matter
scenarios by the condition
$\Lambda_\text{Landau-pole}\gtrsim 10^8\,$~GeV, which implicitly
assumes that the new dynamics does not involve any baryon number
violation.
\subsection{Accidental matter representations}
\label{sec:acc-mat-rep}
The quantum numbers of the new representation are determined by
whether $\boldsymbol{R}$ is a fermion or a scalar. Since
$\boldsymbol{R}$ should preserve the SM symmetries, as pointed out in
\cite{DiLuzio:2015oha} fermionic representations should be such that
operators of the form
\begin{equation}
  \label{eq:effective-operator-DM}
  \mathcal{O}_{N=4}\sim \boldsymbol{R}\,\mathcal{O}_\text{SM}\, ,
\end{equation}
should not be possible writing. Otherwise new
$\mathcal{G}_F=U(3)_Q\otimes U(3)_d\otimes U(3)_u\otimes
U(3)_\ell\otimes U(3)_e$-breaking
sources will be introduced. For scalar representations instead
renormalizable operators are possible without affecting
$\mathcal{G}_F$, provided their quantum numbers do not enable
couplings with SM fermionic bilinears. Bearing in mind this discussion
and that
\begin{equation}
  \label{eq:su2-decompositions}
  \underbrace{\boldsymbol{2}\otimes\dots \otimes 
    \boldsymbol{2}}_{(R-1)\;\;\text{times}}\supset \boldsymbol{R}\, ,
\end{equation}
the first relevant fermionic representation is
$\boldsymbol{R}_F=\boldsymbol{4}_F^Y$, with hypercharge $Y=1/2$ or
$3/2$ determined by whether it couples to $\ell\,H\,H^\dagger$ or
$\ell\,H^\dagger\,H^\dagger$ (here we will use the notation
$\boldsymbol{R}_{F,S}^Y$ where $\boldsymbol{R}$ labels the
representation, $F$ and $S$ its fermionic or scalar character and $Y$
its hypercharge. Hypercharge is normalized according to
$Q=T_3+Y$). The remaining fermionic representations follow from the
rule in~(\ref{eq:su2-decompositions}) with hypercharge fixed by the SM
operator to which they couple. For scalars, at the renormalizable
level, the first representation is indeed a SM singlet,
$\boldsymbol{R}_S=\boldsymbol{1}_S^0$, with the remaining
representations given by $\boldsymbol{R}_S=\boldsymbol{3}_S^0$,
$\boldsymbol{4}_S^{1/2}$, $\boldsymbol{4}_S^{3/2}$ and
$\boldsymbol{6}_S^{1/2}$, with the latter being loop-induced (see
sec. \ref{sec:unstability})\footnote{Note that $SU(2)\otimes U(1)_Y$
  invariance allows for $\boldsymbol{R}_S=\boldsymbol{3}_S^1$. This
  representation however couples to the fermion bilinear
  $\overline{\ell^c}\ell$ and so introduces $\mathcal{G}_F$-breaking
  sources.}. For non-renormalizable operators the first scalar
representation is $\boldsymbol{R}_S=\boldsymbol{2}_S^Y$ with
$Y=3/2, 5/2$. The remaining scalar representations are:
$\boldsymbol{5}_S^Y$ ($Y=0,1,2$), $\boldsymbol{6}_S^Y$ ($Y=3/2, 5/2$),
$\boldsymbol{7}_S^Y$ ($Y=0, 1, 2, 3$), $\boldsymbol{8}_S^Y$
($Y=1/2, 3/2, 5/2, 7/2$) and so on. From this list, viable
representations are selected from cosmological constraints and the
condition of perturbativity.

The neutral component of higher-order $SU(2)$ representations is
cosmologically stable: For scalars
$\boldsymbol{R}_S^Y>\boldsymbol{5}_S^Y$, while for fermions
$\boldsymbol{R}_F^Y\geq\boldsymbol{5}_F^Y$. Direct DM searches
constraints, however, rule out all those representations for which
$Y\neq 0$
\cite{Cirelli:2005uq,Cirelli:2007xd,Cirelli:2009uv}. Perturbativity
criteria as well places constraints on viable representations. Using a
two-loop RGE analysis, ref. \cite{DiLuzio:2015oha} has shown that for
$\boldsymbol{R}_S^0>\boldsymbol{8}_S^0$ and
$\boldsymbol{R}_F^0>\boldsymbol{6}_F^0$ a Landau pole is obtained for
scales below $\sim 10^8\,$~GeV. Thus, in a model defined by SM +
$\boldsymbol{R}$ the accidental matter representations are:
$\boldsymbol{R}_S^Y\leq \boldsymbol{7}_S^0$ and
$\boldsymbol{R}_F^Y< \boldsymbol{5}_F^0$.

A major difference between the \textit{standard} accidental matter
models and the setups we are considering here is that in the latter
there are new degrees of freedom that enter at relatively low scales
($10^6\,$~GeV). The presence of these states induces fast decays of
those representations that otherwise would be cosmologically stable
\cite{Sierra:2016qfa}. Thus they no longer involve a DM particle, and
therefore direct DM constraints no longer hold. This enables $Y\neq 0$
representations, something that is particularly important for the
scalar sextet (see sec. \ref{sec:unstability})\footnote{This argument
  applies as well to $\boldsymbol{5}_F^{1,2}$, however for these
  representations alone a Landau pole is reached at scales below
  $\sim 10^8\,$~GeV. The presence of additional representations at
  $10^6\,$~GeV reduces that scale to values well below those that
  define our perturbativity criteria (see fig. \ref{fig:scales}), and
  so we do not consider them.}.

\begin{table}
  \centering
  \renewcommand{\arraystretch}{1.3}
  \setlength{\tabcolsep}{7pt}
  \begin{tabular}{|c|c|c|c|c|c|c|c|c|c|}
    \hline
    \multicolumn{10}{|c|}{\textbf{Radiative accidental matter representations}}
    \\\hline
    $Y$ & $\boldsymbol{1}_S^Y$ & $\boldsymbol{2}_S^Y$ & $\boldsymbol{3}_S^Y$ 
    & $\boldsymbol{4}_S^Y$ & $\boldsymbol{5}_S^Y$ & $\boldsymbol{6}_S^Y$ 
    & $\boldsymbol{7}_S^Y$ & $\boldsymbol{4}_F^Y$ & $\boldsymbol{5}_F^Y$
    \\\hline
    0 & \cellcolor{LightBlue1}{\ding{51}} & -- & \cellcolor{LightBlue1}{\ding{51}}
    & -- & \cellcolor{LightPink1}{\ding{51}} & -- & \cellcolor{LightPink1}{\ding{53}} 
    & -- & \cellcolor{PaleGreen1}{\ding{51}}
    \\\hline
    1/2 & -- & -- & --
    & \cellcolor{LightBlue1}{\ding{51}} & -- & \cellcolor{LightBlue1}{\ding{51}} 
    & -- & \cellcolor{LightPink1}{\ding{51}} & --
    \\\hline
    1 & -- & -- & --
    & -- & \cellcolor{LightPink1}{\ding{51}} & -- & -- & -- & --
    \\\hline
    3/2 & -- & \cellcolor{LightPink1}{\ding{51}} & --
    & \cellcolor{LightBlue1}{\ding{51}} & -- & \cellcolor{PaleGreen1}{\ding{51}}
    & -- & \cellcolor{LightPink1}{\ding{51}} & --
    \\\hline
    2 & -- & -- & --
    & -- & \cellcolor{LightPink1}{\ding{51}} & -- & -- & -- & --
    \\\hline
    5/2 & -- & \cellcolor{LightBlue1}{\ding{51}} & --
    & -- & -- & \cellcolor{PaleGreen1}{\ding{51}} & -- & -- & --
    \\\hline
  \end{tabular}
  \caption{Accidental matter representations obtained under the assumption
    that $\boldsymbol{R}$ plays an active role in neutrino mass generation.
    Checkmarks (dashes) indicate representations which are (not) viable.
    Blueish cells refer to representations involving renormalizable
    couplings. Reddish (greenish) cells refer instead to representations involving
    dim=5 (dim=6) non-renormalizable couplings. The list differs with what was found in 
    \textit{standard} accidental matter scenarios where lepton number 
    violation occurs at $10^{15}\,$~GeV \cite{DiLuzio:2015oha} in that it
    contains scalar sextets and the septet is not allowed by perturbative criteria.}
  \label{tab:accidental-matter}
\end{table}
In summary, the setups we will consider henceforth are defined by the
accidental matter representations listed in
tab. \ref{tab:accidental-matter}. This list differs from that found in
\textit{standard} accidental matter scenarios in that it contains the
scalar sextet representations, which are enabled due to the
instability of higher-order representations induced by the presence of
additional representations. The UV completed models we will construct
are therefore defined by these representations and subject to the
condition of the full UV model satisfying
$\Lambda_\text{Landau-pole}>10^8\,$~GeV (see
sec. \ref{sec:uv-completions}).
\section{Higher-order $SU(2)$ representations and their cosmological
  instability}
\label{sec:unstability}
As we have already pointed out, higher-order $SU(2)$ representations
in SM + $\boldsymbol{R}$ models are cosmologically stable. The neutral
component of the hypercharge-zero fermion quintet and scalar septet
are---in principle---WIMP DM particles. For these representations tree
level effective operators of the form (operators that induce DM decay)
\begin{equation}
  \label{eq:decay-operator}
  \mathcal{O}_N=\boldsymbol{R}\,\mathcal{O}_\text{SM}\, ,
\end{equation}
with $\mathcal{O}_\text{SM}$ an operator entirely consisting of only
SM fields, are dim=6 and dim=7, respectively. Thus, lifetimes
amounting to $10^{26}\,$~seconds (as required by the non-observation
of $\gamma$-ray, $\nu$, $e^+$ or $p^-$ signals in DM indirect
detection experiments
\cite{Ando:2015qda,Rott:2014kfa,Ibarra:2013zia,Giesen:2015ufa}) are
found for $\Lambda\gtrsim 10^{15}\,$~GeV for $\boldsymbol{5}_F^0$ and
$\Lambda\gtrsim 10^{10}\,$~GeV for $\boldsymbol{7}_S^0$, provided
$m_\text{DM}\subset [5,10]\,$~TeV \cite{Sierra:2016qfa}.  However, if
one considers one-loop induced effective operators one finds that
scalar septet decays are instead driven by dim=5 operators
\cite{DiLuzio:2015oha,DelNobile:2015bqo}, for which not even
$\Lambda=M_\text{Planck}$ leads to sufficiently large DM decay
lifetimes. This observation then singles out the hypercharge-zero
fermion quintet as the only representation containing a viable DM
particle.

It is worth pointing out that this representation is however subject
to stringent constraints coming from indirect DM searches.
Particularly relevant are limits derived from $\gamma$-ray line
searches from the galactic center, for which it has been found that if
the Milky Way exhibits a Navarro-Frenk-White or Einasto DM profile
this representation is not viable either
\cite{Cirelli:2015bda,Garcia-Cely:2015dda}. It can be however
consistently considered in the context of cored profiles such as
Burket or Isothermal. Or by relaxing the hypothesis of WIMP DM,
allowing for $Y=\epsilon\ll 1$ and so leading to millicharged DM
scenarios \cite{DelNobile:2015bqo}.

In scenarios defined by UV completions of the operator in
fig. \ref{fig:dimension-five-LNV-DM}, slow decays of higher-order EW
representations do not hold anymore. The point is that at
$\Lambda=10^6\,$~GeV new states kick in, introducing new
renormalizable couplings that allow writing down operators of type
(\ref{eq:decay-operator}) with cutoff scales fixed by neutrino
data. Roughly one can write $m_\nu\sim v^2Y_\nu^4/16\pi^2\Lambda$,
which means that for $m_\nu=m_\text{Atm}=50\,$meV
\cite{Capozzi:2013csa,Forero:2014bxa,Gonzalez-Garcia:2014bfa} and
$Y_\nu=1$ the cutoff scale should be below $\sim 10^{13}\,$~GeV. This
scale is below the value required for cosmological stability of the
fermion quintet, thus showing that these setups are not reconcilable
with slow DM decays (for a more detailed discussion see
\cite{Sierra:2016qfa,Ahriche:2016rgf}). This can be put more precisely
in the context of explicitly broken symmetries: DM slow decays can be
understood as due to an accidental $\mathbb{Z}_2$ symmetry under which
$\boldsymbol{R}\to -\boldsymbol{R}$ and $X_\text{SM}\to X_\text{SM}$,
and which results as a consequence of the SM gauge symmetry. For
$\boldsymbol{R}=\boldsymbol{5}_F^0$, UV completions of the operator in
fig. \ref{fig:scales} always allow for couplings that break
$\mathbb{Z}_2$, hence DM instability\footnote{Sufficiently small
  $\mathbb{Z}_2$-breaking couplings can lead to cosmological stability
  of the neutral component of the multiplet and so can be used to
  reconcile minimal DM with loop-induced neutrino masses, see
  ref. \cite{Cai:2016jrl} for more details.}.

\begin{figure}[t!]
  \centering
  \includegraphics[scale=0.6]{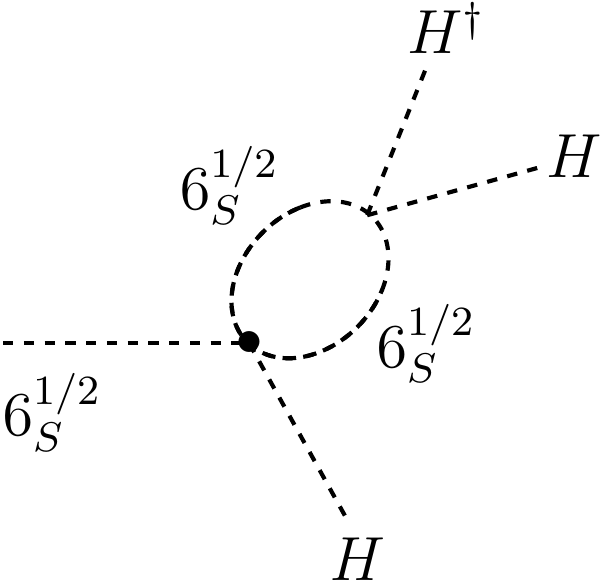}
  \hspace{2cm}
  \includegraphics[scale=0.6]{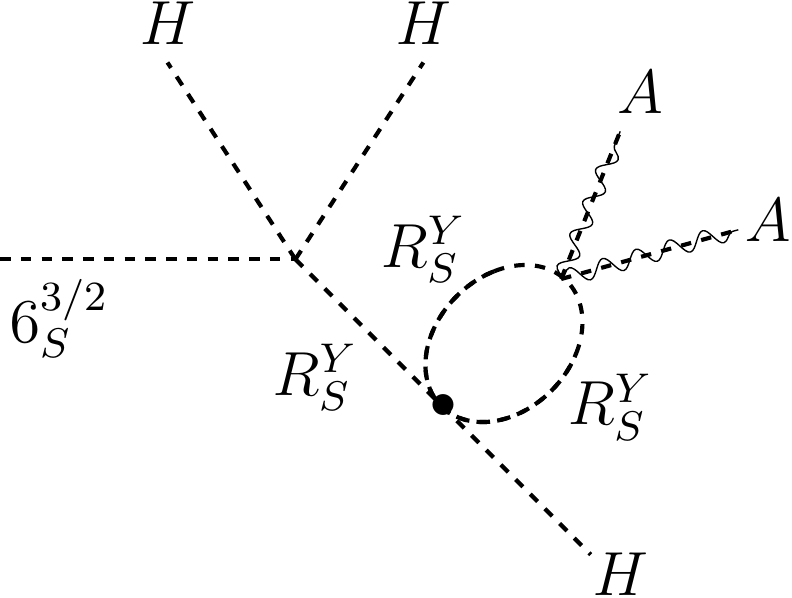}
  \\[2mm]
  \begin{minipage}{1cm}
    $(a)$
  \end{minipage}
  \hspace{4cm}
  \begin{minipage}{1cm}
    $(b)$
  \end{minipage}
  \\[4mm]
  \includegraphics[scale=0.6]{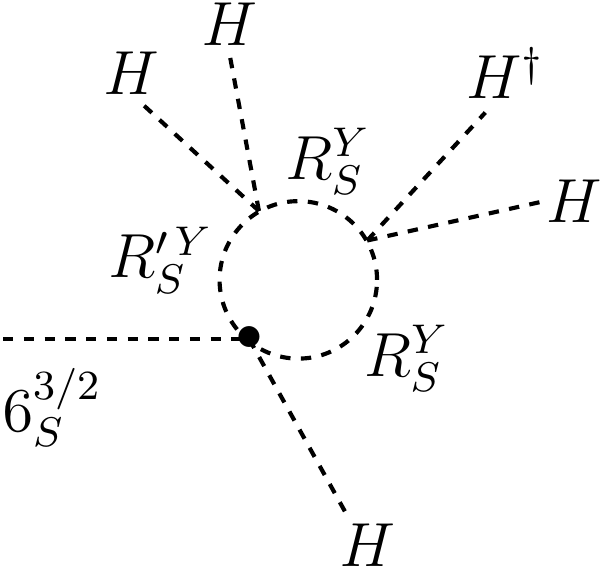}
  \hspace{2cm}
  \includegraphics[scale=0.6]{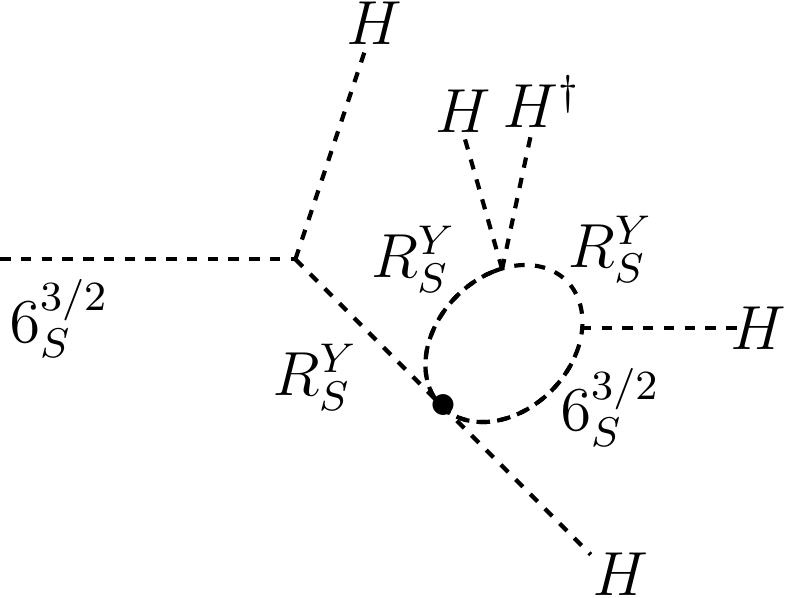}
  \\[2mm]
  \begin{minipage}{1cm}
    $(c)$
  \end{minipage}
  \hspace{4cm}
  \begin{minipage}{1cm}
    $(d)$
  \end{minipage}
  \\[4mm]
  \hspace{-1cm}
  \includegraphics[scale=0.6]{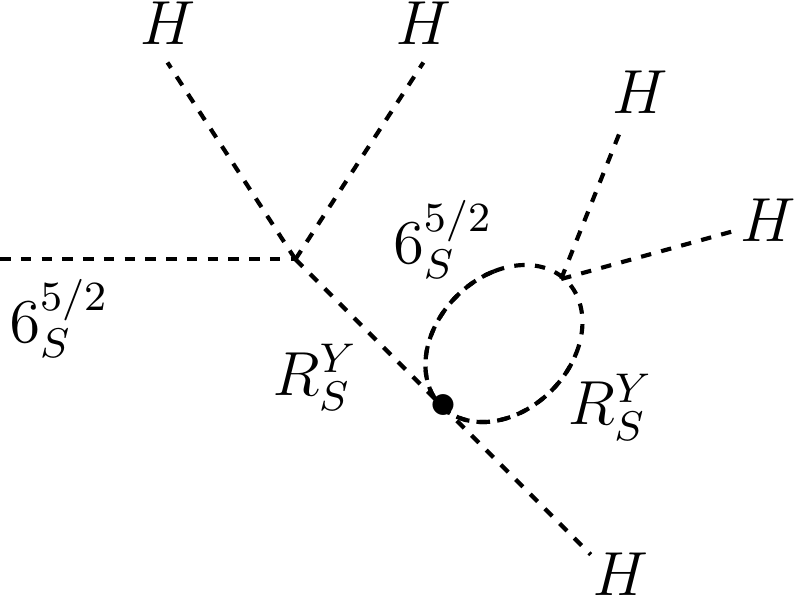}
  \\[2mm]
  \hspace{-1cm}
  \begin{minipage}{1cm}
    $(e)$
  \end{minipage}
  \caption{Loop-induced operators responsible for fast scalar sextet
    decays. Diagram $(a)$ is always present regardless of whether
    there are additional representations. For diagram $(b)$,
    $\boldsymbol{R}_S^Y=\boldsymbol{4}_S^{1/2},
    \boldsymbol{6}_S^{1/2}$.
    The wiggly lines refer to $A=H$ for $\boldsymbol{4}_S^{1/2}$ and
    $A=V$, with $V$ a vector boson, for $\boldsymbol{6}_S^{1/2}$. In
    diagram $(c)$,
    $\{\boldsymbol{R}_S^Y,\boldsymbol{R}_S^{\prime
      Y}\}=\{\boldsymbol{5}_S^1,\boldsymbol{5}_S^2\},
    \{\boldsymbol{4}_S^{5/2},\boldsymbol{6}_S^{3/2}\},
    \{\boldsymbol{6}_S^{5/2},\boldsymbol{6}_S^{3/2}\}$.
    In diagram $(d)$, $\boldsymbol{R}_S^Y=\boldsymbol{5}_S^1$. For
    diagram $(e)$,
    $\boldsymbol{R}_S^Y=\boldsymbol{4}_S^{3/2},
    \boldsymbol{6}_S^{3/2}$.
    The dots indicate $\mathbb{Z}_2$-breaking couplings, which ensure
    fast $\varphi^0\subset \boldsymbol{6}_S^Y$ decays thus making them
    suitable accidental matter representations.}
  \label{fig:sextet-decays}
\end{figure}

These arguments can apply to $Y\neq 0$ representations, depending on
the value of $Y$ and on the specific UV completion. Indeed, they are
responsible for the scalar sextet as a viable accidental matter
representation as we now discuss. For $\boldsymbol{6}_S^{1/2}$, gauge
invariance allows the following scalar coupling:
\begin{equation}
  \label{eq:sextet-one-half}
  V\supset \boldsymbol{6}_S^{1/2}\,\boldsymbol{6}_S^{1/2}\,
  \boldsymbol{6}_S^{-1/2}\,H^{-1/2}\, ,
\end{equation}
that explicitly breaks the accidental $\mathbb{Z}_2$ symmetry. The
presence of this coupling enables the operator in
fig. \ref{fig:sextet-decays}-$(a)$, which induces fast decay processes
of the neutral component of the multiplet,
$\varphi^0\subset \boldsymbol{6}_S^{1/2}$, such as
$\varphi^0\to W^\pm\,W^\mp\,Z^0$. Since slow decays are---in
general---not possible, its density is rapidly depleted and thus
should be consistently included in the list of possible accidental
matter representations. Note that this operator can be written even in
the absence of additional representations, and so this conclusion
proves to be true even in \textit{standard} accidental matter
scenarios.

Operators for $\boldsymbol{6}_S^{3/2}$ can also be written, but in
contrast to the $\boldsymbol{6}_S^{1/2}$ case they require additional
representations. The different decay operators are shown in
fig. \ref{fig:sextet-decays}, diagrams $(b)-(d)$. As can be seen, they
all involve $\mathbb{Z}_2$-breaking couplings and therefore lead to
fast decay processes of the lightest component of the multiplet, in
this case $\varphi^0\subset \boldsymbol{6}_S^{3/2}$. Among those
processes one can identify e.g. $\varphi^0\to W^\pm\,W^\mp\,Z$ or
$\varphi^0\to W^\pm\,W^\mp\,h^0$. These couplings are of three kinds:
independent of $\boldsymbol{6}_S^{3/2}$ and bilinear and linear in
$\boldsymbol{6}_S^{3/2}$. Explicitly, for each operator, they are
given by
\begin{alignat}{2}
  (b):&\qquad \boldsymbol{R}_S^{1/2}\,
  \boldsymbol{R}_S^{1/2}\,\boldsymbol{R}_S^{-1/2}\,H^{-1/2}\, ,&
  \qquad\quad \left(\boldsymbol{R}_S^{1/2}=\boldsymbol{4}_S^{1/2},
  \boldsymbol{6}_S^{1/2}\right)\, ,
  \nonumber\\
  (c):&\qquad \boldsymbol{6}_S^{3/2}\,
  \boldsymbol{6}_S^{3/2}\,\boldsymbol{R}_S^{-5/2}\,H^{-1/2}\, ,&
  \qquad\quad \left(\boldsymbol{R}_S^{-5/2}=\boldsymbol{4}_S^{-5/2},
    \boldsymbol{6}_S^{-5/2}\right)\, ,
  \nonumber\\
  (c):&\qquad \boldsymbol{6}_S^{3/2}\,
  \boldsymbol{5}_S^{-2}\,\boldsymbol{5}_S^{1}\,H^{-1/2}\, ,
  \nonumber\\
  (d):&\qquad \boldsymbol{6}_S^{3/2}\,
  \boldsymbol{5}_S^{-1}\,\boldsymbol{5}_S^{-1}\,H^{1/2}\, .
\end{alignat}
As we will show in sec. \ref{sec:uv-completions}, these couplings
cover all possible UV completions associated with this representation,
apart from one which involves the following representations:
$\boldsymbol{5}_S^{2}$, $\boldsymbol{4}_F^{3/2}$ and
$\boldsymbol{5}_F^{2}$ and for which we did not find a coupling
enabling fast decay ($\mathbb{Z}_2$-breaking coupling). This UV
completion therefore has not been included in our analysis. Finally,
for $\boldsymbol{6}_S^{5/2}$ several decay operators can be written
too. Here, however, we present just a single one (see
fig. \ref{fig:sextet-decays}-$(e)$). The reason is that for this
representation only few UV completions are consistent with our
perturbativity criteria (see sec. \ref{sec:uv-completions}), and this
operator covers all of them. Being $\mathbb{Z}_2$-breaking it induces
fast decays of the neutral component of the multiplet and so allows
for a viable accidental $Y=5/2$ sextet.
\section{Radiative accidental matter}
We now turn to the discussion of UV completions of the operator in
fig. \ref{fig:dimension-five-LNV-DM}. For certain representations, in
particular for higher-order ones, a certain UV completion can as well
generate a dim=7 or dim=9 lepton-number-violating operator. In these
cases one can find therefore regions in parameter space where the
effective neutrino mass matrix receives contributions from several
operators, or even where the neutrino mass matrix is entirely
determined by the higher-order operator. Our assumption here is that
$m_\nu$ is solely determined by the operator in
fig. \ref{fig:dimension-five-LNV-DM}, and this imposes a condition on
the scale of the beyond-the-SM (BSM) degrees of freedom. The
contribution to neutrino masses from the tree level dim=7
lepton-number breaking operator is
$m_\nu^\text{dim=7}\simeq v^4/\Lambda^3$. Thus, when compared with the
one-loop contribution from the Weinberg operator it can be seen that
$\Lambda\gtrsim 3\,$~TeV guarantees
$m_\nu^\text{dim=7}<m_\nu^\text{dim=5}$, a condition satisfied by the
BSM spectrum that defines the radiative accidental matter scenarios we
are discussing (see fig. \ref{fig:scales}).

\subsection{Additional constraints: $\rho$ parameter and BBN}
\label{sec:additional-constraints}
Beyond the constraints we have already mentioned, there are other
constraints one needs to bear in mind. Of particular relevance are
those related with the breaking of the custodial symmetry, which place
bounds on the mass of the accidental matter representations. A
detailed analysis of these constraints has been presented in
\cite{DiLuzio:2015oha} and therefore here we discuss only those
aspects that directly apply to radiative scenarios.  In the limit
$\sin\theta_W\to 0$ ($g^\prime\to 0$), the weak gauge bosons $W^\pm$
and $Z$ transform as a triplet of an $SU(2)_{L+R}$ global symmetry,
which implies $m_{W^\pm}=m_Z$. In that limit the SM $\rho$-parameter,
defined as $\rho=m_W^2/m_Z^2\cos^2\theta_W$, is one. Departures from
this limit removes the gauge bosons mass degeneracy through
$\cos\theta_W$, but still one finds $\rho_\text{tree}=1$, with small
deviations induced by radiative corrections, that remain under control
due to the $SU(2)_{L+R}$ custodial symmetry. This value is consistent
with its experimental value,
$\rho_\text{Exp}=1.0004^{+0.0003}_{-0.0004}$ \cite{Agashe:2014kda}.

Contributions to the $\rho$-parameter from BSM scalar fields that
develop vevs can produce sizeable deviations from such value. For a
set of scalars $\{\varphi_{T,Y}\}$ that acquire a vev,
$\langle \varphi_{T,Y}\rangle$, and whose total weak isospins are $T$
and their hypercharges are $Y$, the tree level $\rho$-parameter reads
\cite{Gunion:1989we}
\begin{equation}
  \label{eq:rho-param}
  \rho_\text{tree}=\frac{\sum_{T,Y}
    c_{T,Y}\left[T(T+1) - Y^2\right]\langle \varphi_{j,Y}\rangle^2}
  {2\sum_Y Y^2\langle \varphi_{j,Y}\rangle^2}\, ,
\end{equation}
where $c_{T,Y}=1$ ($c_{T,Y}=1/2$) for complex (real) fields and
$Q=T_3 + Y$.  There is an infinite set of scalar fields that satisfy
$\rho_\text{tree}=1$, determined by the condition
\begin{equation}
  \label{eq:conditions-rho-eq-one}
  \left(T + \frac{1}{2}\right)^2 - 3 Y^2 = \frac{1}{4}\, .
\end{equation}
The list of the viable scalar representations (that contain a neutral
component) are determined by the following quantum numbers:
$(T,Y)=\{(0,0),(1/2,\pm 1/2), (3,\pm 2), (25/2,\pm 15/2), \cdots\}$
\footnote{The phenomenology of a $(T,Y)=(3,\pm 2)$ state that mixes
  with the SM Higgs doublet has been studied in
  ref. \cite{Hisano:2013sn}.}.  Which shows that apart from the
singlet, none of the other accidental matter scalar representation
satisfies such condition and thus their vevs are subject to
constraints. With only the Higgs and a single extra scalar field,
expression (\ref{eq:rho-param}) at order
$\langle \varphi_{T,Y}\rangle^2/v^2$ can be cast according to
\begin{equation}
  \label{eq:rho-tree-SM-plus-extra-scalar}
  \rho_\text{tree} - 1
  \simeq 2\left\{c_{T,Y}\left[T(T+1) - Y^2\right] - 2Y^2\right\}
  \frac{\langle\varphi_{T,Y}\rangle^2}{v^2}\, ,
\end{equation}
from which using the experimental upper limit for $\rho_\text{Exp}$
one finds $\langle\varphi_{T,Y}\rangle/v\lesssim 1\%$
\cite{DiLuzio:2015oha}. This constraint is particularly important for
accidental matter representations which develop an induced vev, namely
the triplet and the quartet (the singlet does as well but its vev does
not contribute to $\rho_\text{tree}$, as we have mentioned). For these
representations this restriction translates into a lower bound on
their masses, which can be (roughly) estimated from the minimization
condition of the corresponding scalar potentials:
\begin{equation}
  \label{eq:triplet-quartet-pot}
  V_{\boldsymbol{3}}\supset m_3^2\,|\boldsymbol{3}_S^0|^2 
  + \mu\,\boldsymbol{3}_S^0\,H^\dagger\,H\, ,
  \qquad
  V_{\boldsymbol{4}}\supset m_4^2\,|\boldsymbol{4}_S^Y|^2 
  + \lambda\,\boldsymbol{4}_S^Y\,S^3\, ,
\end{equation}
with $S^3=H\,H^\dagger\,H^\dagger$ for $Y=1/2$ and
$S^3=H^\dagger\,H^\dagger\,H^\dagger$ for $Y=3/2$. The result reads
\begin{equation}
  \label{eq:mass-bounds-rho}
  m_3\gtrsim 1568\left(\frac{\mu}{10^2\,\text{GeV}}\right)^{1/2}\,\text{GeV}\, ,
  \qquad
  m_4\gtrsim 246\left(\frac{\lambda}{10^{-2}}\right)^{1/2}\,\text{GeV}\, .
\end{equation}
As can be seen, these values are consistent with radiative accidental
matter models and in particular with the benchmark scenario we have
chosen. It is worth emphasizing that in models where several of these
states are found, these bounds will be more stringent with the values
estimated to increase multiplicatively with the number of states.

Non-vanishing contributions to the $\rho$-parameter arise as well from
radiative corrections to gauge boson masses. Mass splittings between
the different components of a representation $\boldsymbol{R}$ lead to
large radiative contributions, provided the splittings are large
\cite{Veltman:1977kh}. These splittings can arise from one-loop
corrections (for fermions and scalars) and from off-diagonal terms in
the tree level scalar mass matrices. The former are of order MeV
\cite{Cirelli:2005uq} and so lead to negligible corrections to the
$\rho$-parameter. The latter instead can involve large splittings and
so can induce in turn sizeable deviations on $\rho$. However, when
used to derive limits on scalar masses, the values found are less
competitive than those in (\ref{eq:mass-bounds-rho}) or those coming
from direct accelerator searches \cite{DiLuzio:2015oha}.

We now turn to the discussion of the constraints arising from
BBN. Long-lived particles with lifetimes larger than $\sim 0.1$
seconds can significantly affect light-elements abundances through
their electromagnetic and/or hadronic activity. Thus, consistency with
observed light-elements abundances translates into constraints which
lead e.g. to lower bounds on their masses/couplings
\cite{Cyburt:2002uv,Kawasaki:2004qu}. Whether such constraints hold
for the scenarios we are considering here depends on the lifetime of
the different decay processes.  For the representations in
tab. \ref{tab:accidental-matter} there are two types of
decays. Intermultiplet decay processes in which heavier components of
a multiplet undergo decays into lighter components, and decays of the
lightest state (LS) into SM particles. The former are fast decay
processes such as e.g. $R^+\to R^0 + \pi^+$, and so they take place at
early times well before BBN. The latter can be fast too, depending on
whether the lightest particle can decay via renormalizable couplings
or, in case it does not, on the cutoff scale. As it has been discussed
in sec. \ref{sec:acc-mat-rep}, effective decay processes are driven by
either dim=5 or dim=6 operators, for which the decay lifetimes for the
LS can be estimated to be
\begin{align}
  \label{eq:lifetimes-dim5}
  \tau_\text{dim=5}&\simeq 9.8\,
  \left(\frac{\Lambda}{10^6\,\text{GeV}}\right)^2\,
  \left(\frac{10^3\,\text{GeV}}{m_\text{LS}}\right)^3\,\text{fs}\, ,
  \\
  \label{eq:lifetimes-dim6}
  \tau_\text{dim=6}&\simeq 18\,
  \left(\frac{\Lambda}{10^6\,\text{GeV}}\right)^4\,
  \left(\frac{10^3\,\text{GeV}}{m_\text{LS}}\right)^5\,\mu\text{s}\, .
\end{align}
Note that this result assumes that the LS can directly decay via the
non-renormalizable operator. However, this is not the case for
$\boldsymbol{2}_S^{5/2}$ and $\boldsymbol{5}_S^Y$ ($Y=1,2$) which
instead follow cascade decays mediated by off-shell heavier components
of the representation. These processes have been studied in ref.
\cite{DiLuzio:2015oha} assuming $\Lambda=10^{15}\,$~GeV and resulting
in lifetimes amounting to $\sim 10^3\,$~seconds, but the rescaling of
these results according to our cutoff scale lead to lifetimes
comparable to those in (\ref{eq:lifetimes-dim5}). All in all, the
decay processes of the accidental matter representations in radiative
scenarios are fast, with the largest lifetimes amounting to at most
microseconds, and so BBN constraints are of no relevance.
\label{sec:UV-completions}
\begin{figure}
  \centering
  \includegraphics[scale=0.7]{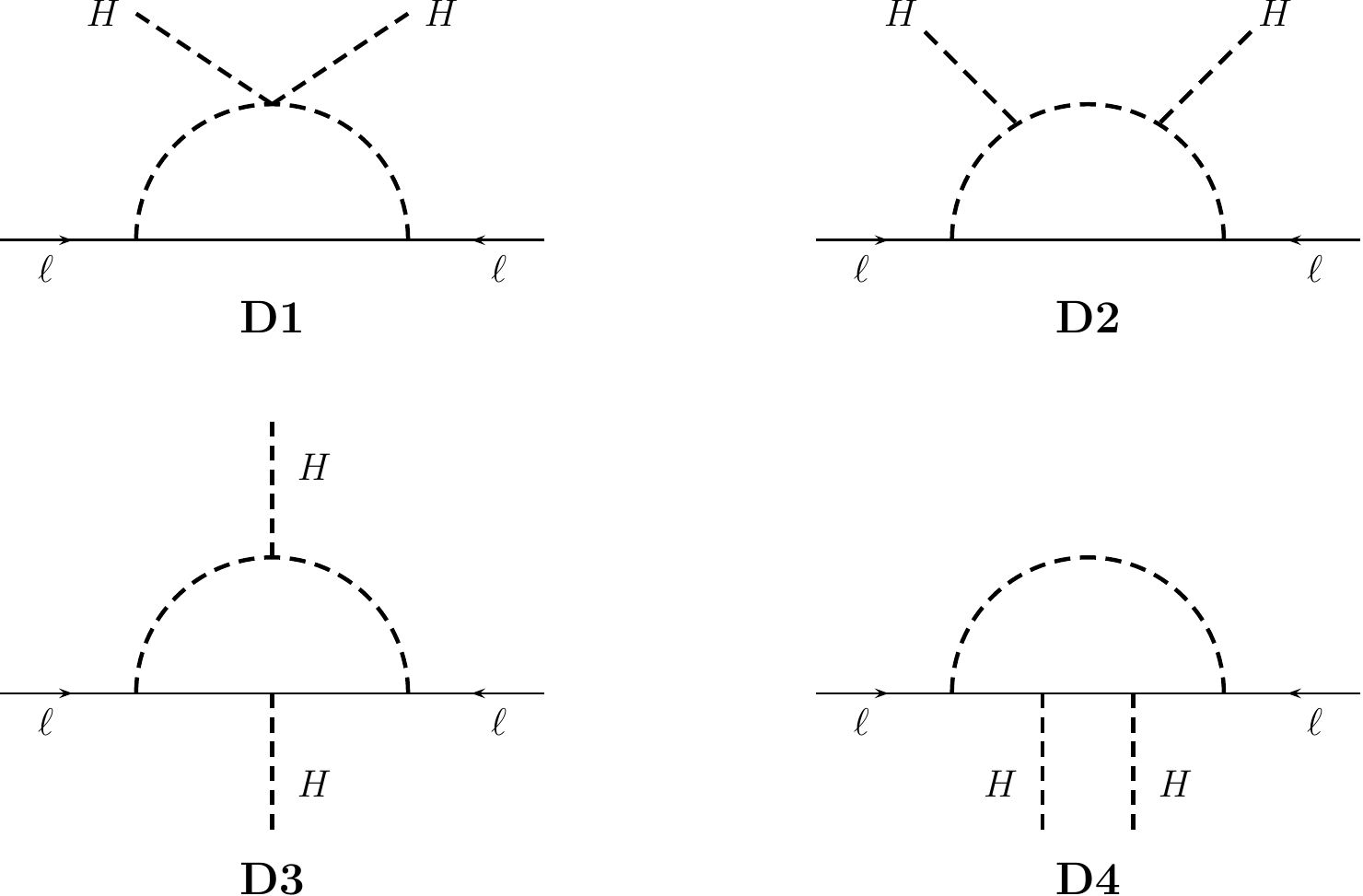}
  \caption{One-loop diagrams for UV realizations of the dim=5 operator
    in fig. \ref{fig:dimension-five-LNV-DM}.}
  \label{fig:one-loop-weinberg}
\end{figure}
\subsection{UV completions and perturbativity}
\label{sec:uv-completions}
Our assumption is that at or above $10^6\,$~GeV new degrees of freedom
defining different UV completions for the operator in
fig. \ref{fig:dimension-five-LNV-DM} become available. For a given
representation the full set of UV completed models can be derived by
considering the diagrams in fig.~\ref{fig:one-loop-weinberg}, which
correspond to all possible irreducible one-loop realizations of the
operator in fig. \ref{fig:dimension-five-LNV-DM} \cite{Bonnet:2012kz}
\footnote{Such classification does exist as well for the two-loop
  case, see ref. \cite{Sierra:2014rxa}.}. We systematically fix the
accidental matter representation within the loop in each of the
diagrams and then fix the $SU(2)\times U(1)_Y$ quantum numbers of the
remaining fields according to
$\boldsymbol{R}\otimes \boldsymbol{2}=\boldsymbol{R}\pm 1$. UV
completions involving hypercharge-zero fermion singlets or triplets or
hypercharge-one scalar triplets are discarded, since they lead to
seesaw neutrino masses (``seesaw filtering criterion''). These cases
are found for $\boldsymbol{1}_S^0$, $\boldsymbol{3}_S^0$ and
$\boldsymbol{4}_F^Y$ ($Y=1/2,3/2$). Rather than explicitly listing the
resulting models, which amount to hundreds and that can be
straightforwardly derived, we list in tab. \ref{tab:representations}
the representations which are needed in each case. These results are
then used to identify those models that lead to the highest Landau
pole scales.

\begin{table}
  \centering
  \renewcommand{\arraystretch}{1.4}
  \setlength{\tabcolsep}{2.5pt}
  \begin{tabular}{|c|ccccc|ccccc|c|c|c|c|}
    \hline
    \multirow{2}{*}{\textbf{A.M.}}
    &
    \multicolumn{5}{c|}{\multirow{2}{*}{\textbf{Fermion sector}}}
    &
    \multicolumn{5}{|c}{\multirow{2}{*}{\textbf{Scalar sector}}}
    &
    \multicolumn{4}{|c|}{\textbf{\# of models}}
    \\\cline{12-15}
    &&&&&&&&&&&\textbf{D1} & \textbf{D2} & \textbf{D3} & \textbf{D4}
    \\\hline
    $\boldsymbol{2}_S^Y$ & $\boldsymbol{1}_F^{Y\pm 1/2}$ 
    & $\boldsymbol{2}_F^Y$ & $\boldsymbol{3}_F^{Y\pm 1/2}$ 
    & $\boldsymbol{4}_F^Y$ & \multicolumn{1}{c|}{\cellcolor{Gray0}{}}
    & $\boldsymbol{1}_S^{Y\pm 1/2}$
    & $\boldsymbol{2}_S^{Y\pm 1}$ & $\boldsymbol{3}_S^{Y\pm 1/2}$ 
    & $\boldsymbol{4}_S^{Y\pm 1}$
    & \multicolumn{1}{c|}{\cellcolor{Gray0}{}} & 12 & 24 & 17 & 10
    \\\hline
    $\boldsymbol{3}_S^0$ & $\boldsymbol{2}_F^{1/2}$ 
    & $\boldsymbol{4}_F^{1/2}$ & $\boldsymbol{5}_F^0$ 
    & \multicolumn{2}{c|}{\cellcolor{Gray0}{}}
    & $\boldsymbol{1}_S^1$ 
    & $\boldsymbol{2}_S^{1/2}$ & $\boldsymbol{4}_S^{1/2}$ 
    & $\boldsymbol{5}_S^1$
    & \multicolumn{1}{c|}{\cellcolor{Gray0}{}} & 2 & 3 & 1 & 1
    \\\hline
    $\boldsymbol{4}_S^Y$ & $\boldsymbol{2}_F^Y$ 
    & $\boldsymbol{3}_F^{Y\pm 1/2}$ & $\boldsymbol{4}_F^Y$ 
    & $\boldsymbol{5}_F^{Y\pm 1/2}$ & $\boldsymbol{6}_F^Y$
    & $\boldsymbol{2}_S^{Y\pm 1}$ 
    & $\boldsymbol{3}_S^{Y\pm 1/2}$ & $\boldsymbol{4}_S^{Y\pm 1}$ 
    & $\boldsymbol{5}_S^{Y\pm 1/2}$
    & $\boldsymbol{6}_S^{Y\pm 1}$ & 14 & 21 & 15 & 8
    \\\hline
    $\boldsymbol{5}_S^Y$ & $\boldsymbol{3}_F^Y$ 
    & $\boldsymbol{4}_F^{Y\pm 1/2}$ & $\boldsymbol{5}_F^Y$ 
    & $\boldsymbol{6}_F^{Y\pm 1/2}$ & $\boldsymbol{7}_F^Y$
    & $\boldsymbol{3}_S^{Y\pm 1}$ 
    & $\boldsymbol{4}_S^{Y\pm 1/2}$ & $\boldsymbol{5}_S^{Y\pm 1}$ 
    & $\boldsymbol{6}_S^{Y\pm 1/2}$
    & $\boldsymbol{7}_S^{Y\pm 1}$ & 11 & 21 & 13 & 8
    \\\hline
    $\boldsymbol{6}_S^Y$ & $\boldsymbol{4}_F^Y$ 
    & $\boldsymbol{5}_F^{Y\pm 1/2}$ & $\boldsymbol{6}_F^Y$ 
    & $\boldsymbol{7}_F^{Y - 1/2}$ & \multicolumn{1}{c|}{\cellcolor{Gray0}{}}
    & $\boldsymbol{4}_S^Y$ 
    & $\boldsymbol{4}_S^{Y\pm 1}$ & $\boldsymbol{5}_S^{Y\pm 1/2}$ 
    & $\boldsymbol{6}_S^{Y\pm 1}$
    & $\boldsymbol{7}_S^{Y - 1/2}$ & 9 & 15 & 5 & 1
    \\\hline
    $\boldsymbol{4}_F^Y$ & $\boldsymbol{3}_F^{Y\pm 1/2}$ 
    & $\boldsymbol{4}_F^{Y\pm 1}$ & $\boldsymbol{5}_F^{Y\pm 1/2}$ 
    & $\boldsymbol{6}_F^{Y- 1}$ & \multicolumn{1}{c|}{\cellcolor{Gray0}{}}
    & $\boldsymbol{2}_S^Y$ 
    & $\boldsymbol{3}_S^{Y\pm 1/2}$ & $\boldsymbol{4}_S^Y$ 
    & $\boldsymbol{5}_S^{Y\pm 1/2}$
    & $\boldsymbol{6}_S^Y$ & 4 & 5 & 13 & 19
    \\\hline
    $\boldsymbol{5}_F^0$ & $\boldsymbol{3}_F^1$ 
    & $\boldsymbol{4}_F^{1/2}$ & $\boldsymbol{5}_F^1$ 
    & $\boldsymbol{6}_F^{1/2}$ & \multicolumn{1}{c|}{\cellcolor{Gray0}{}}
    & $\boldsymbol{3}_S^0$ 
    & $\boldsymbol{4}_S^{1/2}$ & $\boldsymbol{5}_S^0$ 
    & $\boldsymbol{6}_S^{1/2}$
    & $\boldsymbol{7}_S^0$ & 3 & 5 & 5 & 7
    \\\hline
  \end{tabular}
  \caption{Representations needed for the construction of UV completions of the 
    operator in fig. \ref{fig:dimension-five-LNV-DM} for the different accidental
    matter representations in tab. \ref{tab:accidental-matter}. The last four
    columns refer to the number of viable models determined by the UV completions
    in fig. \ref{fig:one-loop-weinberg} and selected according to the condition
    that the particle content of the model does not enable type-I, type-II
    or type-III seesaw and that $\alpha_{1,2}$ remain perturbative at least up
    to $10^8\,$~GeV.}
  \label{tab:representations}
\end{table}
For $\boldsymbol{1}_S^0$ all viable models are forbidden by the
``seesaw filtering criterion''.
For $\boldsymbol{3}_S^0$ some models are found, but still the filtering
criterion removes most of them leaving just few. For
$\boldsymbol{4}_F^Y$, instead, this criterion removes only few
models. Instead, in this case, a fairly large number of such models
are found to be non-viable because they lead to non-perturbative
effects below $10^8\,$~GeV, something found as well for other
higher-order $SU(2)$ representations.

Models that become non-perturbative below $10^8\,$~GeV, as defined by
our benchmark scenario, are identified by using two-loop RGEs subject
to the following energy thresholds (see appendix
\ref{sec:explicit-UV-completions}):
\begin{itemize}
\item From $M_Z$ and up to $m_{\boldsymbol{R}}$ ($\boldsymbol{R}$
  being the accidental matter representation), that we take to be
  $1\,$~TeV, the particle content is entirely given by the SM.
\item From $m_{\boldsymbol{R}}$ and up to $10^6\,$~GeV, where according to
  our benchmark scenario the UV completions for the operator in
  fig.~\ref{fig:dimension-five-LNV-DM} kick in, the particle content
  is determined by the SM + $\boldsymbol{R}$.
\item Above $10^6\,$~GeV, RGE running takes into account the SM +
  $\boldsymbol{R}$ + $\boldsymbol{R}^\prime$, with
  $\boldsymbol{R}^\prime$ referring to representations that define the
  UV completions.
\end{itemize}
For low-order $SU(2)$ representations up to the triplet, Landau poles
are found at rather high scales, ranging from $10^{11}\,$~GeV up to
$10^{19}\,$~GeV, with the exception being few models for
$\boldsymbol{2}_S^{5/2}$ for which $\alpha_1=g_1^2/4\pi$ develops a
Landau pole at $\sim 10^8\,$~GeV, due to the large hypercharges of the
extra representations. Thus, apart from this representation, all
low-order radiative accidental matter models are consistent with
perturbativity up to $10^8\,$~GeV.

\begin{figure}[t!]
  \centering
  \includegraphics[scale=0.53]{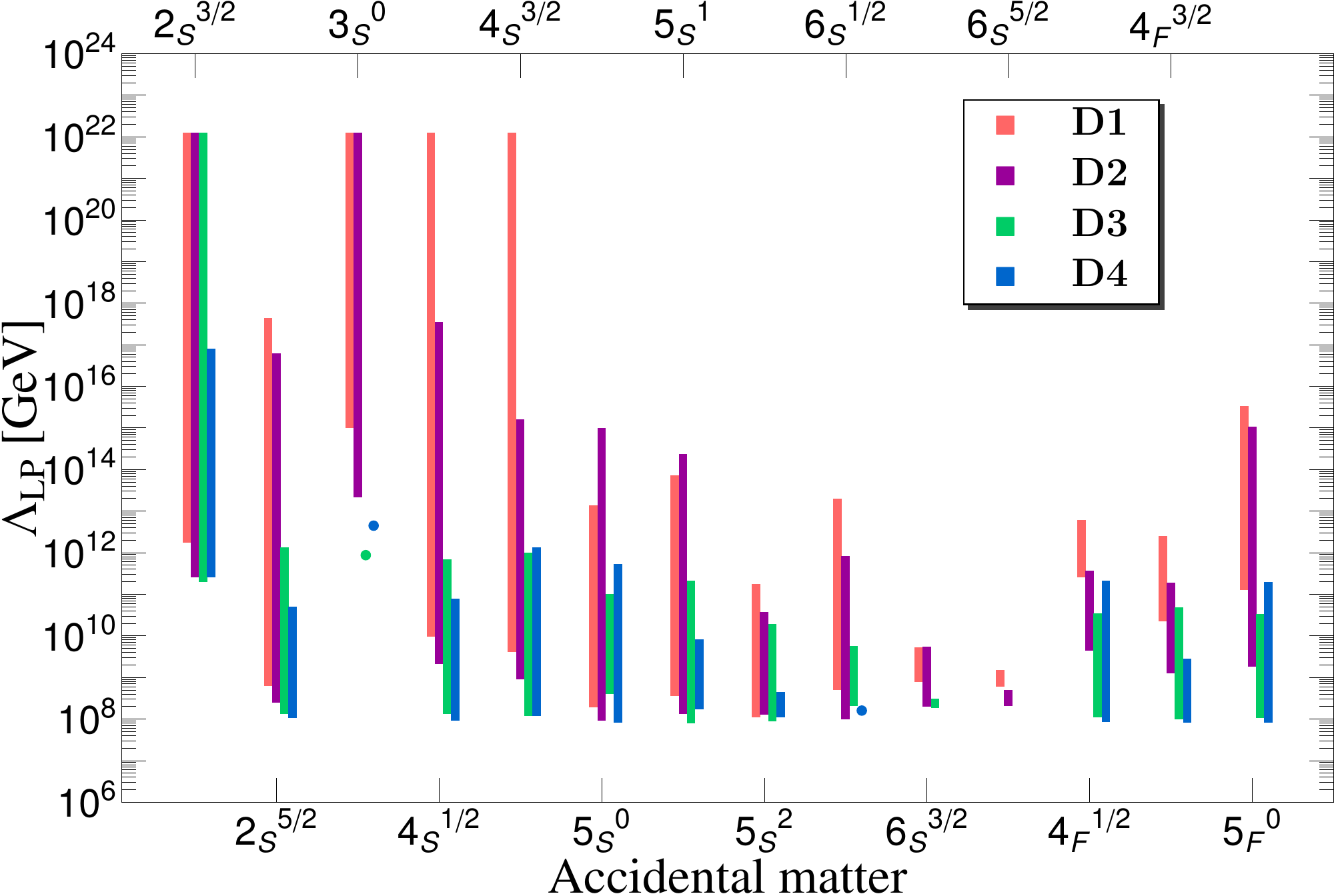}
  \caption{Range of the Landau pole scales for the UV models of the
    operator in fig. \ref{fig:dimension-five-LNV-DM}, displayed
    according to the category models defined by diagrams
    \textbf{D1}-\textbf{D4} in fig. \ref{fig:one-loop-weinberg}. Bars
    exceeding $10^{19}\,$~GeV indicate that the Landau pole is reached
    above $M_\text{Planck}$, and do not refer to any precise value.
    For some representations/models only a point is displayed: For
    $\boldsymbol{3}_S^0$ a single model (\textbf{D3} and \textbf{D4})
    is possible after the ``seesaw filtering criterion'' is applied,
    for $\boldsymbol{6}_S^{1/2}$ (\textbf{D4}) and
    $\boldsymbol{6}_S^{3/2}$ (\textbf{D3}) only a single model reaches
    a Landau pole above $10^8\,$~GeV. Furthermore, for
    $\boldsymbol{6}_S^{3/2}$ all models in category \textbf{D4} reach
    a Landau pole below $10^8\,$~GeV, while for
    $\boldsymbol{6}_S^{5/2}$ all models in \textbf{D3} and \textbf{D4}
    do so.}
  \label{fig:landau-pole}
\end{figure}
For higher-order accidental matter representations this behavior
remains like that for the $\boldsymbol{4}_S^{1/2}$ and also for
$\boldsymbol{4}_S^{3/2}$, but for $\boldsymbol{4}_S^{3/2}$ some models
fail to pass the perturbativity condition. For $\boldsymbol{5}_S^0$
all models are consistent with perturbativity and
$\Lambda_\text{LP}^\text{highest}$ remains at $10^{13}\,$~GeV,
depending on the model category (defined by diagrams
\textbf{D1}-\textbf{D2}). For $\boldsymbol{5}_S^Y$ ($Y=1,2$), the
largest Landau pole scales are somehow degraded with values even as
low as $10^9\,$~GeV for $\boldsymbol{5}_S^2$, again depending on the
model category. In these cases many models reach Landau poles well
below $10^8\,$~GeV, and so fail to pass the pertubativity cut. This
trend persists for the remaining accidental matter representations,
including the $\boldsymbol{4}_F^Y$, being very pronounced for
$\boldsymbol{6}_S^{3/2}$ and $\boldsymbol{6}_S^{5/2}$ for which there
are model categories that do not contain any model satisfying the
perturbativity criterion. These results are summarized in
fig.~\ref{fig:landau-pole}, where we have plotted the range for the
Landau pole scale for the different accidental matter representations
in each category model.

Among all the viable models we select those for which the Landau pole
scale is the largest. These radiative accidental matter models are
arguably the most compelling ones. For models involving
$\boldsymbol{3}_S^0$ and $\boldsymbol{2}_S^{3/2}$ several setups for
which $\Lambda_\text{LP}\gtrsim 10^{19}\,$~GeV are found. In these
cases the selection criterion is that of the model involving the least
number of representations. In almost all cases the corresponding
models are \textbf{D1}-based, something somehow expected since these
UV completions involve the least number of fermions and so gauge
couplings run slower. In all cases as well we have found that
$\alpha_2$ reaches the Landau pole before $\alpha_1$ does, with a
single exception given by $\boldsymbol{2}_S^{5/2}$. The models are
shown in tab.~\ref{tab:compelling}, where it can be seen that models
with $\boldsymbol{5}_S^2$, $\boldsymbol{6}_S^{3/2}$ and
$\boldsymbol{6}_S^{5/2}$ are somehow disfavored by the relatively low
Landau pole scale.
\begin{table}[t!]
  \centering
  \renewcommand{\arraystretch}{1.4}
  \setlength{\tabcolsep}{5pt}
  \begin{tabular}{|c|c|c|c|}
    \hline
    \textbf{Acc. Matter} & \textbf{Extra representations}& \textbf{Model}
    &$\Lambda_\text{LP}$\\\hline
    %
    $\boldsymbol{2}^{3/2}_S$ & $\boldsymbol{1}^1_F$ \hspace{1cm} 
    $\boldsymbol{2}^{1/2}_S$ & \textbf{D1} & $>10^{19}\,$~GeV\\\hline
    $\boldsymbol{2}^{5/2}_S$ & $\boldsymbol{1}^2_F$ \hspace{1cm} 
    $\boldsymbol{2}_S^{3/2}$ & \textbf{D1} & $3.41\times 10^{17}\,$~GeV\\\hline
    $\boldsymbol{3}^0_S$ & $\boldsymbol{2}^{1/2}_F$ \hspace{1cm} 
    $\boldsymbol{1}^1_S$ & \textbf{D1} & $>10^{19}\,$~GeV\\\hline
    $\boldsymbol{4}^{1/2}_S$ & $\boldsymbol{3}^1_F$ \hspace{1cm} 
    $\boldsymbol{2}^{3/2}_S$ & \textbf{D1} & $>10^{19}\,$~GeV\\\hline
    $\boldsymbol{4}^{3/2}_S$ & $\boldsymbol{3}^1_F$ \hspace{1cm} 
    $\boldsymbol{2}^{1/2}_S$ & \textbf{D1} & $>10^{19}\,$~GeV\\\hline
    $\boldsymbol{5}^0_S$ & $\boldsymbol{5}^0_F$ \hspace{1cm} 
    $\boldsymbol{4}^{1/2}_S$ & \textbf{D2} & $7.94\times 10^{14}\,$~GeV\\\hline
    $\boldsymbol{5}^1_S$ & $\boldsymbol{4}^{1/2}_F$ \hspace{1cm} 
    $\boldsymbol{3}^0_S$ & \textbf{D2} & $5.74\times 10^{13}\,$~GeV\\\hline
    \cellcolor{Snow2}{$\boldsymbol{5}^2_S$} & \cellcolor{Snow2}{$\boldsymbol{6}^{3/2}_F$} \hspace{1cm} 
    $\boldsymbol{5}^1_S$ & \cellcolor{Snow2}{\textbf{D1}} & \cellcolor{Snow2}{$1.03\times 10^9\,$~GeV}\\\hline
    $\boldsymbol{6}^{1/2}_S$ & $\boldsymbol{5}^0_F$  & 
    \textbf{D1} & $1.54\times 10^{13}\,$~GeV\\\hline
    \cellcolor{Snow2}{$\boldsymbol{6}^{3/2}_S$} & \cellcolor{Snow2}{$\boldsymbol{5}^1_F$} \hspace{1cm} 
    \cellcolor{Snow2}{$\boldsymbol{4}^{1/2}_S$} & \cellcolor{Snow2}{\textbf{D1}} & \cellcolor{Snow2}{$4.19\times 10^9\,$~GeV}\\\hline
    \cellcolor{Snow2}{$\boldsymbol{6}^{5/2}_S$} & \cellcolor{Snow2}{$\boldsymbol{5}^2_F$} \hspace{1cm} 
    $\boldsymbol{6}^{3/2}_S$ & \cellcolor{Snow2}{\textbf{D1}} & \cellcolor{Snow2}{$7.4\times 10^8\,$~GeV}\\\hline
    $\boldsymbol{4}^{1/2}_F$ & $\boldsymbol{3}^0_S$ \hspace{1cm} 
    $\boldsymbol{5}^1_S$ & \textbf{D1} & $4.76\times 10^{12}\,$~GeV\\\hline
    $\boldsymbol{4}^{3/2}_F$ & $\boldsymbol{3}^2_S$ \hspace{1cm} 
    $\boldsymbol{5}^1_S$ & \textbf{D1} & $1.95\times 10^{12}\,$~GeV\\\hline
    $\boldsymbol{5}^0_F$ & $\boldsymbol{4}^{1/2}_S$ & 
    \textbf{D1} & $2.63\times 10^{15}\,$~GeV\\\hline
  \end{tabular}
  \caption{Radiative accidental matter models for which the Landau pole is 
    reached at the highest possible scale. Apart from
    $\boldsymbol{5}_S^0$ and $\boldsymbol{5}_S^1$, all 
    models are \textbf{D1}-based. This list therefore 
    defines the most \textit{compelling} radiative accidental 
    matter models. Note that the relatively low Landau pole scale for 
    $\boldsymbol{5}_S^2$, $\boldsymbol{6}_S^{3/2}$ and
    $\boldsymbol{6}_S^{5/2}$ disfavor these models.}
  \label{tab:compelling}
\end{table}
\subsection{Lepton flavor violation: generic approach}
\label{sec:LFV}
In this section we quantify the expected size of SM charged-lepton
flavor violating (CLFV) radiative processes in the models depicted in
tab.~\ref{tab:compelling}\footnote{CLFV in models with higher-order
  $SU(2)$ representations, which are among the models shown in
  tab.~\ref{tab:compelling}, have been considered in
  e.g. refs. \cite{Cai:2016jrl,Cai:2011qr,Chowdhury:2015sla}.}. Three
body CLFV decay processes, in particular $\mu^+ \to e^+e^-e^+$, and
$\mu-e$ conversion in nuclei are relevant as well due to the large
sensitivity of near-future experimental facilities: $Mu3e$ at PSI aims
at measuring $\mu^+ \to e^+e^-e^+$ to a precision of $10^{-16}$
\cite{Bravar:2015vja,Berger:2014vba}, while PRISM/PRIME at J-PARC
$\mu-e$ conversion in nuclei down to $10^{-18}$
\cite{Kuno:2005mm,Barlow:2011zza}. Results for these processes will be
presented elsewhere \cite{diego}.

Rather than sticking to a particular realization or analyzing them all
we make use of the fact that the problem follows a generic
treatment. Since most of the compelling models are determined by
\textbf{D1} diagrams, we will focus on such models for which a
schematic Lagrangian can be used for the discussion:
\begin{equation}
  \label{eq:Lag-schem}
  \mathcal{L} = \overline{\ell^c}_i\,Y_{i\alpha}\,P_L\,F_\alpha\,S_a 
  + \lambda_{ab}\,S_a\,S_b\,H\,H
  + \overline{F}_\alpha\,Y_{j\alpha}P_L\,\ell_j\,S_b
  + \text{H.c.}\, ,
\end{equation}
where $F$ ($S$) refer to vectorlike fermions (scalars) in any of the
representations displayed in tab.~\ref{tab:compelling}, and so $SU(2)$
contractions are assumed. Sum over lepton flavor and fermion
generation indices is understood, while $a$ and $b$ just label
different scalars. Note that in addition to the terms in
(\ref{eq:Lag-schem}), there are also pure scalar and fermion mass
terms (that can be of Majorana type if $Y(F)=0$) which we are not
writing, but are essential since they determine scalar mixing and
eventually fermion mixing too. Furthermore, we are assuming for
simplicity that Yukawa couplings are the same regardless of the scalar
to which the fermion bilinear is coupled.

We take two generations of vectorlike fermions, which is the minimal
number required to generate two non-zero light neutrino masses in this
simplified setup we have assumed. This is a direct consequence of
assuming same Yukawa couplings, for different Yukawa couplings a
single fermion will suffice. Note that in models where more than a
single fermion is needed the Landau pole scale will be below the
numbers quoted in tab.~\ref{tab:compelling}.  Since in this section we
aim just at showing that CLFV effects are within reach and we are not
specifying quantum numbers we stick to the values in
tab.~\ref{tab:compelling}.  Components of the scalar multiplets (that
we assume there is only one copy per representation) with the same
electric charge $Q$ mix through the scalar coupling in
(\ref{eq:Lag-schem}). Depending on $Q$, their mass eigenstates will
contribute to the neutrino mass operator. One can distinguish for
example: $(i)$ one pair of scalar mass eigenstates with electric
charge $Q$ determine the neutrino mass matrix, $(ii)$ two pairs with
electric charges $Q$ and $Q^\prime$ are responsible for the
operator. Just to mention a couple of cases, in the model for
$\boldsymbol{2}_S^{3/2}$, states
$\varphi^+\subset \boldsymbol{2}_S^{3/2}$ and
$\rho^+\subset \boldsymbol{2}_S^{1/2}$ mix, and their mass eigenstates
$(S_1^+,S_2^+)$ lead to two diagrams that determine the neutrino mass
matrix. In contrast, in case $\boldsymbol{4}_S^{1/2}$, states
$\varphi^+\subset \boldsymbol{2}_S^{3/2}$ and
$\rho^+\subset \boldsymbol{4}_S^{1/2}$ and
$\varphi^{++}\subset \boldsymbol{2}_S^{3/2}$ and
$\rho^{++}\subset \boldsymbol{4}_S^{1/2}$ mix and define the
eigenstates $(S_1^+,S_2^+)$ and $(S_1^{++},S_2^{++})$ that in turn
lead to the neutrino mass matrix.

Therefore, after rotation to the scalar mass eigenstate bases and
regardless of the radiative accidental matter model, the neutrino mass
matrix consist of a series of diagrams that guarantee cancellation of
the divergent piece of the Passarino-Veltman (PV) function
$B_0(0,m_{S_a}^2,m_F)$ \cite{Passarino:1978jh}. The finite piece
defines in turn the neutrino mass matrix that reads:
\begin{equation}
  \label{eq:nmm-generic-case}
  \left(m_\nu\right)_{ij}=\frac{1}{16\pi^2}\sum_{\alpha=1,2} m_{F_\alpha}\,Y_{i\alpha}Y_{j\alpha}\,
  \sum_a\,\Theta^2_a\,B_0^\text{fin}(0,m_{S_a}^2,m_{F_\alpha}^2)\, ,
\end{equation}
where $\Theta_a^2$ parameterizes scalar mixing and satisfies
$\sum_a \Theta_a^2=0$. In the simplest case $a=1,2$,
$\Theta_a^2=(-)^a\sin\theta\cos\theta$ ($\theta$ the mixing angle) and
so the finite piece of the PV function under the sum over $a$ can be
written according to
\begin{equation}
  \label{eq:pv-zero}
  B_0^\text{fin}(0,m_{S_a}^2,m_{F_\alpha}^2)\to 
  B_0^\text{fin}(0,m_{S_a}^2,m_{F_\alpha}^2)=\frac{m_{S_a}^2}{m_{S_a}^2 - m_{F_\alpha}^2}
  \,\log\left(\frac{m_{S_a}^2}{m_{F_\alpha}^2}\right)\, .
\end{equation}
Note that the neutrino mass matrix is rank-two and so there is a
massless light neutrino, as already anticipated. To determine the
Yukawa structure required by neutrino oscillation data
\cite{Forero:2014bxa,Capozzi:2013csa,Gonzalez-Garcia:2014bfa}, that
combined with the mass scale for the fermions and scalars determines
the LFV rates, we start by recasting the neutrino mass matrix
according to
\begin{equation}
  \label{eq:nmm-recasted}
  \boldsymbol{m_\nu}= \boldsymbol{Y}\cdot \widehat{\boldsymbol{F}}\cdot \boldsymbol{Y}^T\, ,
\end{equation}
where $\boldsymbol{Y}$ is a $3\times 2$ Yukawa coupling matrix whose
entries are given by $Y_{i\alpha}$ and $\widehat{\boldsymbol{F}}$ is a
dimensionful $2\times 2$ diagonal matrix whose non-vanishing entries read:
\begin{equation}
  \label{eq:diagonal-matrix-nmm}
  \widehat{\boldsymbol{F}}_\alpha=-\frac{\Theta^2\,m_{F_\alpha}}{16\pi^2}
  \left[
    \frac{m_{S_1}^2}{m_{S_1}^2-m_{F_\alpha}^2}\log\left(\frac{m_{S_1}^2}{m_{F_\alpha}^2}\right)
    -
    \frac{m_{S_2}^2}{m_{S_2}^2-m_{F_\alpha}^2}\log\left(\frac{m_{S_2}^2}{m_{F_\alpha}^2}\right)
  \right]\, .
\end{equation}
With the mass matrix written as in (\ref{eq:nmm-recasted}) the Yukawas
can be then parameterized \`a la Casas-Ibarra \cite{Casas:2001sr},
namely
\begin{equation}
  \label{eq:casas-ibarra}
  \boldsymbol{Y}=\boldsymbol{U}^*\,\boldsymbol{\widehat{m}_\nu}^{1/2}\,\boldsymbol{R}^T\,
  \widehat{\boldsymbol{F}}^{-1/2}\, .
\end{equation}
Here $\boldsymbol{U}$ is the lepton mixing matrix and $\boldsymbol{R}$
is a $3\times 2$ orthogonal complex matrix that can be written as
\cite{Ibarra:2003up}
\begin{equation}
  \label{eq:casas-ibarra-2-times-3}
  \boldsymbol{R}=
  \begin{pmatrix}
    0 & \cos z & \sin z\\
    0 & -\sin z & \cos z\\
  \end{pmatrix}\, ,
\end{equation}
where $z$ is a complex angle. With the aid of
eq. (\ref{eq:casas-ibarra}), radiative CLFV processes can be calculated
by fixing the scalar mass spectrum and using neutrino data.

The branching fractions for these processes can be written according
to \cite{Lavoura:2003xp}
\begin{equation}
  \label{eq:lfv-brs}
  \text{Br}(l_i\to l_j\gamma)=
  \frac{\Gamma(l_i\to l_j\,\gamma)}{\Gamma_\text{Tot}^{l_i}}=
  \frac{m_i^5}{4096\,\pi^5\,\Gamma_\text{Tot}^{l_i}}
  \left|
    \sum_{a,\alpha=1}^2
  \frac{Y_{i\alpha}Y_{j\alpha}^*}{m_{S_a}^2}
  \left[
    Q_F\,G_1(t_{\alpha a})
    +
    (Q_F+1)\,G_2(t_{\alpha a})
  \right]
  \right|^2\, ,
\end{equation}
with $t_{\alpha a}=m_{F_\alpha}^2/m_{S_a}^2$,
$\Gamma_\text{Tot}^{l_i}$ the total decay width of $l_i$ and the loop
functions given by
\begin{align}
  \label{eq:loop-fun}
  G_1(t)=\frac{2 + 3t - 6t^2 + t^3 + 6\,\log t}{12(t-1)^4}
  \qquad
  \text{and}
  \qquad
  G_2(t)=\frac{1 - 6t - 3t^2 + 2t^3 - 6t^2\,\log t}{12(t-1)^4}\, .
\end{align}
With these results at hand we calculated the muon and tau decay
branching ratios as a function of the heaviest fermion mass. For that
aim we fixed fermion masses according to $m_{F_2}-m_{F_1}=500\,$~GeV,
while randomly varying $m_{F_2}$ in the range $[10^4,10^9]\,$~GeV. The
scalar spectrum according to $m_{S_2}=10^3\,$~GeV and
$m_{S_1}=500\,$~GeV, neutrino low-energy observables to their best-fit
point values \cite{Forero:2014bxa} and scalar mixing in the range
$\Theta^2=[10^{-12},10^{-8}]$, randomly varied too. For simplicity we
have taken $z$ real and equal to $\pi/10$. We have checked that the
result is pretty insensitive to the value of $z$ as long as $z$ is
real. The result as well is not very sensitive to the value of $Q_F$
provided $Q_F<2$. For all points in the scan we have checked
max$(\boldsymbol{Y})<\sqrt{4\pi}$.  Note that this parameter choice
has nothing special and has been taken just to exemplify the typical
CLFV behavior one expects in these scenarios.

\begin{figure}
  \centering
  \includegraphics[scale=0.6]{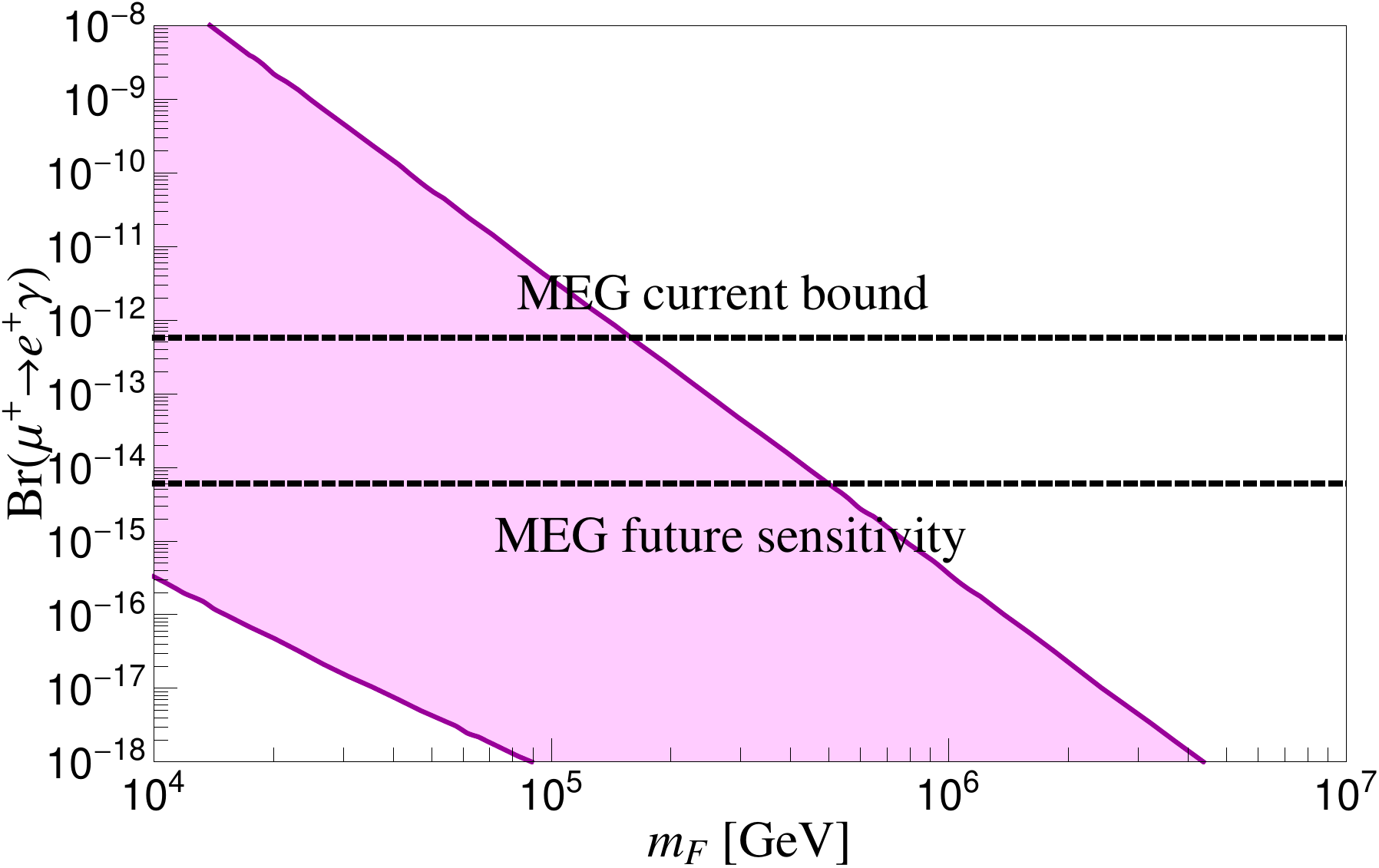}
  \caption{Decay branching fraction for $\mu\to e\gamma$ as a function
    of a generic fermion mass $m_F$. The upper horizontal line
    indicates current MEG bound \cite{Adam:2013mnn}, whereas the lower
    line MEG future sensitivity \cite{Cheng:2013paa}.  This result has
    been derived by assuming that the process is mediated by a pair of
    charged fermions (scalars) with $Q_F=+1$ ($Q_S=+2$). The result is
    representative of what is expected from the radiative accidental
    matter models listed in tab.~\ref{tab:compelling} (compelling
    radiative accidental matter models), with possible numerical
    variations determined by the specific model and/or parameter space
    region.}
  \label{fig:mu-to-e-gamma}
\end{figure}
Fig. \ref{fig:mu-to-e-gamma} shows the result for $\mu\to e \gamma$ as
a function of the heavy fermion mass. It shows that for a reasonable
mass range this process falls within the region determined by MEG
near-future sensitivities \cite{Cheng:2013paa}. The result is
representative of what is expected for the radiative accidental matter
models listed in tab. \ref{tab:compelling}. Since it has been done
without sticking to a particular realization, it is of course subject
to numerical variations determined by the details of the model and its
particular behavior with relevant parameters. For radiative tau decay
processes we have found that $\tau \to e\gamma$ lies one order of
magnitude below SuperKEKB sensitivities \cite{Aushev:2010bq}, while
$\tau\to \mu\gamma$ can be within the range of observability if the
mass gap between $\boldsymbol{R}$ (the accidental matter
representation) and the radiative dim=5 operator UV completion is not
large, $\mathcal{O}(m_\text{UV})\sim 10\,m_{\boldsymbol{R}}$. This,
however, would lead to a lower $\Lambda_\text{LP}$ scale and so
according to our approach is disfavored.
\section{Conclusions}
\label{sec:concl}
Accidental matter models involve BSM degrees of freedom that
automatically preserve the SM accidental and approximate symmetries up
to a certain cutoff scale. Assuming that this scale is determined by
lepton number violation and in the absence of a ``non-conventional''
suppression mechanism, this scale is fixed at $\sim 10^{15}\,$~GeV. In
this paper, we studied accidental matter models assuming that the
accidental matter representations play an active role in neutrino mass
generation (\textit{radiative accidental matter} models), something
that we have shown necessary leads to radiatively induced neutrino
masses and therefore to a suppressed LNV scale that can naturally be
as low as $10^6\,$~GeV.

By defining a benchmark \textit{radiative accidental matter} scenario
(see fig. \ref{fig:scales}), we have shown that in this new context
higher-order accidental matter representations are no longer
cosmologically stable. In particular, we have shown that this
observation combined with a lower perturbative scale enables
$Y=1/2,3/2,5/2$ scalar sextets. By considering EW and BBN constraints,
we have as well derived lower bounds on the masses of various
accidental matter representations, showing that masses of about
$1\,$~TeV always imply a consistent picture.

We have identified the different UV completions of the radiative
dimension five operator in fig. \ref{fig:dimension-five-LNV-DM}.  We
have systematically studied their perturbative behavior and filtered
viable models according to whether $\alpha_1$ and $\alpha_2$ remain
perturbative up to at least $10^8\,$~GeV, our results are summarized
in fig. \ref{fig:landau-pole}. From this classification we have
determined the most compelling \textit{radiative accidental matter}
models by the condition that the Landau pole scale is the largest
possible. Our results are listed in tab.~\ref{tab:compelling}. For the
resulting models we have studied in a fairly model-independent way the
typical size of radiative CLFV effects. According to our findings,
$\mu\to e\gamma$ is within reach for all models provided the UV
completion of the operator in fig. \ref{fig:dimension-five-LNV-DM}
does not exceed $\sim 10^6\,$~GeV. In contrast, for
$\tau\to \mu\gamma$ observability requires a somewhat BSM compressed
spectrum that in turn leads to a lower Landau pole scale, something
disfavored by our approach but not excluded. Thus, these setups offer
a rich LFV phenomenology that increases their experimental testability
and which motivates further phenomenological studies of the resulting
setups \cite{diego}.
\section{Acknowledgments}
We would like to thank 
Luca di Luzio for useful comments.  This
work was supported by the ``Fonds de la Recherche Scientifique-FNRS''
under grant number 4.4501.15. The work of C.S. is supported by the
``Universit\'e de Li\`ege'' and the EU in the context of the
MSCA-COFUND-BeIPD project.
\appendix
\section{Two-loop Renormalization group equations}
\label{sec:explicit-UV-completions}
The evolution of the gauge coupling constants,
$\alpha_{i}\,(i=1,2,3)$, with the energy scale $\mu$, at two-loop
level, is given by the Renormalization Group Equations~(RGEs)
\begin{equation}
\label{eq:RGE_2loop}
\begin{aligned}
  \mu\frac{d}{d\mu}\alpha^{-1}_i&=
  -\frac{b_{i}}{2\pi} - \frac{1}{8\pi^2} \sum_j b_{ij}\,\alpha_j\,,
\end{aligned}
\end{equation}
with $\alpha_1=5/3\alpha_y$, $\alpha_i=g^2_i/4\pi$ and $b_i$ and
$b_{ij}$ the one- and two-loop beta functions, respectively. For a
generic multiplet with
$G_\text{SM}=SU(3)_C\times SU(2)_L\times U(1)_Y$ quantum numbers
$(d_3, d_2, y)$, $b_i$ and $b_{ij}$ are given
by~\cite{Jones:1981we,Machacek:1983tz}
\begin{equation}
\label{eq:beta_1loop}
b_i=\begin{cases}
  -\frac{11}{3}\, C_i,& \text{for gauge bosons} \\[2mm]
  \,\frac{4}{3}\,f\, T_i\,d_k\,d_m,& \text{for fermions} \\[2mm]
  \,\frac{1}{3}\,s\, T_i\,d_k\,d_m,& \text{for scalars} 
\end{cases}
\end{equation}
and 
\begin{equation}
\label{eq:beta_2loop}
b_{ij}=\begin{cases}
  -\frac{34}{3}\, C^2_i\,\delta_{ij},& \text{for gauge bosons}\\[2mm]
  f\,\left[4\,C_i+\frac{20}{3}\,C_i(\text{Adj})\right]\,T_i\,d_k\,d_m\,\delta_{ij}\,
  +\left[\,4\,f\,C_j\,T_i\,d_k\,d_m\right]_{i\neq j},& \text{for fermions} \\[2mm]
  s\,\left[4\,C_i+\frac{2}{3}\,C_i(\text{Adj})\right]\,T_i\,d_k\,d_m\,\delta_{ij}\,
  +\left[\,4\,s\,C_j\,T_i\,d_k\,d_m\right]_{i\neq j},& \text{for scalars} 
\end{cases}
\end{equation}
with $s=1/2$ ($s=1$) for real (complex) scalars, while $f=1/2$ ($f=1$)
for Weyl (Dirac) fermions. $d_k$ and $d_m$ are the dimensions of the
multiplet with respect to the remaining subgroups, $G_k$ and $G_m$,
and $m,k\neq i$. The quantities $C_i$ and $T_i$ are respectively the
Casimir invariant and the Dynkin index for the multiplet under
consideration with respect to the subgroup $G_i\subset G_\text{SM}$.
The Casimir for the adjoint representation of $G_i$ is denoted by
$C_i(\text{Adj})$. The Casimir invariants and Dynkin indices ($C_2$
and $T_2$) for $SU(2)$ representations up to octets are given in
tab.~\ref{tab:invariants}. For $U(1)_Y$ one has instead
$T_1=y^2$. Note that since all the multiplets we consider are color
singlets they do not contribute to $SU(3)$ running and hence we do not
specify neither $C_3$ nor $T_3$. For the gauge couplings at
$M_Z=91.188$~GeV we use~\cite{Agashe:2014kda}:
\begin{equation}
\label{eq:data_at_Mz}
\alpha_1(M_Z)=0.016923 \,,\quad
\alpha_2(M_Z)=0.03374 \,,\quad
\alpha_3(M_Z)=0.1173  \,
\end{equation}
and the SM beta functions, $b^\text{SM}_i$ and $b^\text{SM}_{ij}$ which read 
\begin{equation}
\label{eq:bSM}
b^\text{SM}_i=\left(\frac{41}{10},-\frac{19}{6},-7\right)\,,\qquad 
b^\text{SM}_{ij}=\begin{pmatrix}
\frac{199}{50} & \frac{27}{10} & \frac{44}{5} \\[2mm]	
 \frac{9}{10} & \frac{35}{6} & 12 \\[2mm]
 \frac{11}{10} & \frac{9}{2} & -26 	
\end{pmatrix}.
\end{equation}
\begin{table}[t]
  \renewcommand{\arraystretch}{1.4}
  \setlength{\tabcolsep}{5pt}
  \begin{center}
    \begin{tabular}{c|cccccccc}
      \hline
      & $\boldsymbol{1}$ & $\boldsymbol{2}$ & $\boldsymbol{3}$
      & $\boldsymbol{4}$ & $\boldsymbol{5}$ & $\boldsymbol{6}$
      & $\boldsymbol{7}$ & $\boldsymbol{8}$  \\
      \hline\\[-2ex]
      $T_2$ & 0 &  1/2  &  2  &   5    & 10  & 35/2  & 28 & 42 \\[1ex]
      $C_2$ & 0 &  3/4  &  2  &  15/4  &  6  & 35/4 &  12 & 63/4 \\
      \hline
    \end{tabular}
  \end{center}
  \caption{Dynkin index ($T_2$) and Casimir invariant ($C_2$) for $SU(2)$ 
    representations up to dimension eight.}
  \label{tab:invariants}
\end{table}
%
\bibliography{references}
\end{document}